\documentclass[manuscript,screen,authorversion,nonacm]{acmart}
\AtBeginDocument{%
  }

\usepackage{graphicx}
\usepackage{booktabs}
\usepackage{multirow}
\usepackage{enumitem}
\usepackage{amsmath,amsfonts}
\usepackage{xurl}
\usepackage{hyperref}
\usepackage{array}
\usepackage{pifont}
\newcommand{\cmark}{\ding{51}}
\newcommand{\xmark}{\ding{55}}
\usepackage[ruled,vlined]{algorithm2e}

\begin{document}

\title[Technical Report of FlowGRN]{FlowGRN: Scalable and Dropout-Robust Gene Regulatory Network Inference via Flow Matching-Based Trajectory Reconstruction (Technical Report)}

\author{Tsz Pan TONG}
\email{tszpan.tong@uni.lu}
\orcid{0000-0001-8111-5886}
\author{Jun PANG}
\email{jun.pang@uni.lu}
\orcid{0000-0002-4521-4112}
\affiliation{%
  \institution{Department of Computer Science, University of Luxembourg}
  \city{Esch-sur-Alzette}
  \country{Luxembourg}
}
\affiliation{%
  \institution{Institute for Advanced Studies, University of Luxembourg}
  \city{Esch-sur-Alzette}
  \country{Luxembourg}
}

\renewcommand{\shortauthors}{Tong and Pang}

\begin{abstract}
Inferring gene regulatory networks (GRNs) from single-cell RNA sequencing (scRNA-seq) data offers insights into cellular behavior, but is complicated by the lack of temporal information and the prevalence of dropout noise.
To address these challenges, we present FlowGRN, a method that integrates conditional flow matching and score matching for robust trajectory reconstruction with dynGENIE3 for scalable GRN inference.
FlowGRN incorporates a novel cell similarity measure that is resilient to dropout effects in high-dimensional scRNA-seq data.
Evaluation on the BEELINE benchmark demonstrates that FlowGRN achieves state-of-the-art performance on both synthetic and experimental datasets.
Ablation studies validate the importance of both the dropout-robust similarity measure and the trajectory reconstruction step, highlighting FlowGRN's ability to accurately model dynamic regulatory relationships.
\end{abstract}




\maketitle

\section{Introduction}
Gene regulatory networks (GRNs), which map the intricate interplay between genes and their regulators, are fundamental to our understanding of cellular life.
Accurate deciphering of these networks is essential for elucidating cellular processes and for designing therapeutic strategies against complex diseases~\cite{van2018integrative}.
Recent advances in single-cell RNA sequencing (scRNA-seq) technology have enabled the profiling of gene expression at the single-cell level, providing a wealth of data and valuable insights into cellular dynamics.
However, accurately inferring GRNs from scRNA-seq data remains challenging, primarily because sequencing data are snapshots of cellular states, lacking explicit dynamic information, and dealing with extremely high-dimensional gene spaces.

Current scRNA-seq technology requires stopping the cell evolution process to capture snapshots of cell states.
To balance sequencing depth and resolution, the sampling time points are often sparse, and thus key dynamic changes such as cell transitions are unobserved.
Thus, cell developmental trajectories are lost during sequencing, and the low temporal resolution further hinders our ability to restore cellular dynamics.
In addition, GRN inference involves tens of thousands of genes.
At this scale, the Euclidean distance becomes less informative due to the curse of dimensionality, and the scalability of inference methods becomes a critical issue.

Facing the challenges of loss of cell dynamics and high dimensionality, researchers have taken various approaches to infer GRNs from scRNA-seq data.
Given that gene regulation involves interactions between macromolecules governed by the laws of physics and chemistry, ordinary differential equations (ODEs) have become a natural choice for modeling dynamics.
Different methods based on ODEs, both statistical~\cite{matsumoto2017scode,aubin2020grisli} and deep learning-based~\cite{tong2020trajectorynet,sha2024tigon}, have been proposed to model cellular dynamics.
However, these methods either oversimplify the system or are computationally intractable because of the high dimensionality.
Additionally, restoring dynamics relies on accurate pseudotime estimations for each cell.
However, the performance of different pseudotime estimation algorithms varies on the same dataset~\cite{saelens2019comparison}, and many GRN inference methods~\cite{papili2018sincerities,qiu2020scribe} are highly sensitive to pseudotime accuracy~\cite{pratapa2020benchmarking}.

Another research track maximizes scalability by discarding temporal information and treating the scRNA-seq data as a static snapshot.
GRN inference methods such as GENIE3~\cite{huynh2010genie3}, GRNBOOST2~\cite{moerman2019grnboost2} and PIDC~\cite{chan2017pidc} are widely used and show promising results on various benchmarks~\cite{pratapa2020benchmarking,chen2018evaluating}.
They infer regulatory relationships by quantifying the information gain of each gene in predicting other genes, or the unique information decomposed from the mutual information between genes.
Although these methods are powerful, scalable, and supported by statistical theories, they discard temporal information and make the direction of regulations unidentifiable~\cite{simcha2013identification}.
The loss of dynamic information and the high-dimensionality problem constitute a dilemma in GRN inference, and the community is urging the development of a precise, scalable, and identifiable GRN inference model that utilizes temporal information and reconstructs cellular dynamics in the high-dimensional gene space.

Recently, conditional flow matching (CFM)~\cite{albergo2024multimarginal} has been proposed to use a neural network to model the vector field that flows high-dimensional data from one distribution to another.
Compared to traditional neural ODEs~\cite{tong2020trajectorynet,sha2024tigon}, CFM can learn the flow in >10,000 dimensions efficiently by making a specific assumption about the distribution of pairings between the source and target distributions.
This motivated us to utilize CFM to reconstruct cellular trajectories and infer the GRN from scRNA-seq data.

In practice, we noticed that the heavy dropout noise in scRNA-seq data introduces additional uncertainty to the measurement of the distance between cells, which is a cornerstone in dynamic modeling.
Dropouts are a common phenomenon in scRNA-seq data, where a gene is not detected in a cell even though it is expressed~\cite{jiang2022statistics}.
Some imputation methods, such as MAGIC~\cite{van2018magic}, SAVER~\cite{huang2018saver}, and DCA~\cite{eraslan2019single}, have been proposed to restore missing gene expression from its neighbors and latent data geometry.
Unfortunately, they bring additional false positive edges to the inferred GRN, thus the imputed data are not suitable for GRN inference~\cite{ly2022effect}.
To address dropout noise, we propose a new similarity measure that is robust to dropout in high-dimensional spaces.

In this work, we utilize advances in conditional flow matching and score matching~\cite{tong2024simulation} to reconstruct cellular trajectories.
We then employ dynGENIE3~\cite{huynh2018dyngenie3}, a dynamic version of GENIE3, to infer an identifiable GRN from the reconstructed cellular trajectories with thousands of genes.
We propose a new similarity measure to address dropout noise, which experimentally shows robustness to dropout.
By combining advances in CFM for dynamics modeling, reliable dynGENIE3 for GRN inference, and a dropout-robust cell similarity measure, our scalable GRN inference solution, FlowGRN, outperforms state-of-the-art methods for directed GRN inference in terms of precision.
Our contribution is summarized as follows:
\begin{itemize}
  \item We propose a new cell similarity measure that shows robustness to dropout and the curse of dimensionality.
  \item We apply this new measure on conditional flow matching and show that it can reconstruct cellular trajectories with non-linear dynamics, scalable up to 1,783 genes.
  \item We show that the gradient of the ODEs cannot accurately infer GRNs because of indirect regulations, thus dynGENIE3 is used to infer the GRN from the learned trajectory.
  \item Our method, FlowGRN, has the highest average precision rank in the BEELINE benchmark~\cite{pratapa2020benchmarking} on experimental datasets.
\end{itemize}
Our method is publicly available at \url{https://github.com/1250326/FlowGRN}.

\section{Related Work}
\label{sec:related-work}

\subsection{Gene regulatory network inference}
\label{ssec:grn-inference}

ScRNA-seq data are the primary source for GRN inference, providing snapshots of cell states in the form of a series of gene expression profiles.
To ensure sufficient sequencing depth and resolution, sampling time points are often sparse, which limits the application of time-series models.
Correlation-based methods (ppcor~\cite{kim2015ppcor}, LEAP~\cite{specht2017leap}), information theory-based methods (ARACNe~\cite{margolin2006aracne}, CLR~\cite{faith2007clr}, PIDC~\cite{chan2017pidc}, Scribe~\cite{qiu2020scribe}) and regression methods (TIGRESS~\cite{haury2012tigress}, SINCERITIES~\cite{papili2018sincerities}, GRNVBEM~\cite{sanchez2018grnvbem}) provide a statistically sound approach to infer GRNs from limited data.
However, these methods are bound by the development and strong assumptions of statistical theories.
ODE-based methods (SCODE~\cite{matsumoto2017scode}, GRISLI~\cite{aubin2020grisli}) attempt to model cell dynamics with linear ODEs, which oversimplify system complexity and are computationally inefficient on large GRNs.

Taking advantage of the rapid development of machine learning, ensemble methods (GENIE3~\cite{huynh2010genie3}, dynGENIE3~\cite{huynh2018dyngenie3}, GRNBOOST2~\cite{moerman2019grnboost2}) offer powerful and scalable options for GRN inference.
GENIE3 and GRNBOOST2 are the most recognized and widely used for GRN inference, and the latter is the default option in the SCENIC~\cite{aibar2017scenic} and SCENIC+~\cite{bravo2023scenic+} protocols for coarse selection of regulatory pairs.
Both methods train a tree ensemble model to predict the expression of each target gene from other genes, and use the importance score in each model to quantify the regulatory relationships.
Jump3~\cite{huynh2015jump3} and dynGENIE3 aim to incorporate temporal information, but their reliance on continuous cellular trajectories, which are unobservable in destructive snapshot-based scRNA-seq, has largely limited their application to simulated datasets rather than real-world scenarios.
Recently, researchers have tried to borrow concepts from deep learning, such as convolutional neural networks (DeepDRIM~\cite{chen2021deepdrim}, 3DCEMA~\cite{fan2021gene}, DELAY~\cite{reagor2023delay}), variational autoencoder (DeepSEM~\cite{shu2021deepsem}), and explainable AI (LRP~\cite{keyl2023lrp}).
However, many of these emerging deep learning methods have not yet matched ensemble methods in terms of accuracy and scalability, and thus their practical application is not yet widespread.

GENIE3 and its variants have dominated the field of GRN inference since 2010, and researchers have sought breakthroughs from multi-omics data.
Representative methods include CellOracle~\cite{kamimoto2023celloracle}, Inferelator 3.0~\cite{skok2022high} and Pando~\cite{fleck2023inferring}.
These methods integrate scRNA-seq data with scATAC-seq and spatial transcriptomics data to infer GRNs.
Interested readers can refer to recent reviews~\cite{badia2023gene,loers2024single} for more details.
Although multi-omics data offer new perspectives for GRN inference, revolutionizing our understanding of cellular dynamics using only scRNA-seq data remains an important long-term challenge.

\subsection{The curse of dimensionality}
\label{ssec:dimensionality}
A full-genome GRN inference task contains approximately 20,000 protein-encoding genes.
If we further consider protein-protein interactions, the network size would be even larger.
Thus, scalability is a critical challenge in GRN inference.
Methods such as SCODE~\cite{matsumoto2017scode} and TIGON~\cite{sha2024tigon} use linear mapping and variational autoencoder (VAE) to compress the high-dimensional gene space into an actionable latent space.
Latent space is a powerful tool for dimensionality reduction, and it is widely used in scRNA-seq data visualization and analysis~\cite{mcinnes2018umap,moon2019visualizing}.
DCA~\cite{eraslan2019single} also uses latent space to impute dropout values.
However, compressing gene expression into a lower-dimensional latent space can obscure individual gene interactions.
This can lead to difficulties in accurately reconstructing gene-specific dynamics and potentially introduce spurious relationships when mapping latent interactions back to the gene level.

Fortunately, GRN is sparse, and only a small fraction of genes are regulated by each other.
This enables feature selection and stability selection approaches to tackle high dimensionality.
In brief, feature selection methods select a subset of genes from the whole genome and use them to train models, while stability selection identifies important genes that consistently contribute to the feature-selected models in multiple runs.
Methods like TIGRESS~\cite{haury2012tigress}, LEAP~\cite{specht2017leap}, and GRISLI~\cite{aubin2020grisli} combine two selection methods and show promising results on different benchmarks~\cite{pratapa2020benchmarking,wang2024benchmarking}.
Notably, the popular GENIE3~\cite{huynh2010genie3} also uses a similar approach in general, because its underlying algorithm is random forest, which uses multiple subsets of genes to train decision trees, and the importance score of each gene is averaged across all trees.
This suggests that the ideas of feature and stability selection, along with their variants, are widely applicable in GRN inference.

\subsection{Technical noise challenges in scRNA-seq data}
\label{ssec:dropout}
10x Genomics sequencing technology offers a high-throughput and cost-effective solution for capturing transcripts in individual cells.
It uses Gel Beads-in-Emulsion (GEM) to encapsulate cells and barcodes and count transcripts in each cell~\cite{zheng2017massively}.
This technology is widely used in the field of single-cell biology, but it brings additional dropouts to the gene expression matrix.

Jiang et al.~\cite{jiang2022statistics} classified zero reads from sequencing into biological, technical, and sampling zeros.
Biological zeros are caused by silent genes and uncaptured transcriptional bursts, technical zeros are caused by inefficient reverse transcription from mRNA to cDNA, and sampling zeros are caused by inefficient amplification and low sampling depth.
Thus, only biological zeros and nonzero reads are informative for GRN inference, while technical zeros and sampling zeros are recognized as dropouts.
However, most mainstream regression-based~\cite{huynh2010genie3,haury2012tigress,matsumoto2017scode,huynh2018dyngenie3,moerman2019grnboost2,aubin2020grisli} and correlation-based~\cite{kim2015ppcor,specht2017leap} methods simply use Euclidean distance and do not adequately account for dropout effects.
This introduces biases in modeling, which affect the GRN inference accuracy.

\section{Preliminaries}
\label{sec:preliminaries}

\subsection{Notations and problem definition}
\label{ssec:notations}
Denote $X=\{X^{(t_0)},X^{(t_1)},\cdots,X^{(t_{n-1})}\}$ be a scRNA-seq dataset containing a series of $n$ gene expression profiles, each profile $X^{(t_i)}\in\mathbb{R}^{g\times c_i}$ contains $g$ genes and $c_i$ cells sampled at time $t_i$.
Denote the total number of cells as $c=\sum_{i=0}^{n-1}c_i$.
A cell $x\in\mathbb{R}^{g}$ can be viewed as a point in the $g$-dimensional gene space.
A GRN is a directed graph $\mathcal{G}=(\mathcal{V},\mathcal{E},\omega)$, where $\mathcal{V}$ is the set of genes, $\mathcal{E}\subset \mathcal{V}\times\mathcal{V}$ is the set of edges, and $\omega:\mathcal{E}\rightarrow\mathbb{R}$ is the weight function (regulation strength).
Positive (Negative) weights indicate activation (inhibition) regulations.
An adjacency matrix $A\in\mathbb{R}^{g\times g}$ is a matrix representation of the GRN, where $A_{i,j}=\omega(i,j)$ if $(i,j)\in\mathcal{E}$ and $A_{i,j}=0$ otherwise.
We aim to infer the directed GRN $\mathcal{G}$ from the scRNA-seq dataset $X$ without any prior knowledge of the GRN structure.

\subsection{Conditional flow matching and score matching}
\label{ssec:CFM}
Cellular trajectories are crucial for inferring gene regulatory networks (GRNs).
In a parallel field, structure inference, researchers have shown that trajectory data alone can accurately reconstruct the underlying dynamics of a system~\cite{wang2022isidg,wang2023rcsi,wang2024benchmarking}.
This motivates us to reconstruct the cellular trajectories to facilitate GRN inference.

Neural ODE is a powerful framework for modeling dynamical systems: it learns a vector field whose ODE solution matches observed time-series data.
Recent methods, such as TrajectoryNet~\cite{tong2020trajectorynet} and TIGON~\cite{sha2024tigon}, have successfully applied neural ODEs to scRNA-seq data.
However, their reliance on ODE integration makes them computationally expensive and numerically unstable, limiting their applications to small gene sets (< 50) or low-dimensional latent spaces.
To overcome these limitations, we adopted the [SF]\textsuperscript{2}M model~\cite{tong2024simulation}, an integration-free conditional flow matching and score matching model.
[SF]\textsuperscript{2}M leverages optimal transport (OT) plans between snapshots to define target vector fields, guiding the learning of the dynamics driving cellular transitions.

Conditional flow matching (CFM) trains the vector field referenced to the displacement between the two distributions, whereas neural ODEs train the vector field referenced to the observed trajectories, thereby incurring costly ODE integration.
Viewing scRNA-seq snapshots as a sequence of gene-expression distributions, CFM is a natural choice for learning the vector field driving those flows.

Score matching also trains a vector field, but refers to the score of the distributions of the cell trajectories, restricting the divergence of the CFM model to observed regions.
Score matching can be viewed as the valley in the Waddington landscape, where the vector field points toward regions of higher probability.
In practice, score matching is used as regularization to the CFM model, and the cell dynamics is captured by the CFM model.

In a theoretical sense, [SF]\textsuperscript{2}M can be seen as solving a Schr\"odinger bridge problem
whose marginal evolution of densities is governed by the Fokker-Planck equation.
They model cellular dynamics by the stochastic differential equation (SDE) with the form $dx=u_t(x)dt+g(t)dB_t$, where $u_t(x)$ is the drift and $dB_t$ is the Brownian motion.
This SDE, together with the initial distribution $p_0$, induces probability paths $p_t$ via the continuity equation $\partial_t p_t = -\nabla\cdot(u_t p_t)$.
[SF]\textsuperscript{2}M decomposes the drift $u_t$ into two components: probability flow drift $u_t^\circ=u_t-\frac{1}{2}g^2(t)\nabla \log p_t$ and score $\frac{1}{2}g^2(t)\nabla \log p_t$, which points to regions of higher probability.

Thus, the dynamics can be learned by minimizing two neural networks $v_\theta, s_\theta$ with the following loss function:
\begin{equation}
  \label{eq:loss-unconditional}
  \mathcal{L}_{\text{unconditional}}=\mathbb{E}_{t\sim\mathcal{U}(0,1),x\sim p_t(x)}\left[\left\|v_\theta(x,t)-u_t^\circ(x)\right\|^2 + \lambda(t)^2 \left\|s_\theta(x,t)-\nabla \log p_t(x)\right\|^2\right],
\end{equation}
where $\frac{1}{2}g(t)^2$ is absorbed into the positive weight $\lambda(t)^2$.
$v_\theta$ is called the flow matching model, and $s_\theta$ is the score matching model.

Since $p_t$ is intractable, [SF]\textsuperscript{2}M conditioned on the OT plan $\pi$ to guide the flow.
For a given pair of distributions $p_0$ and $p_1$, the OT plan $\pi$ is a joint distribution with marginal distributions $p_0,p_1$, which minimizes the cost function $\mathbb{E}_{(x_0,x_1)\sim\pi}[c(x_0,x_1)]$.
Sampling $(x_0,x_1)\sim\pi$ allows us to compute:
\begin{align}
  \label{eq:CFM}
  u_t^\circ(x|x_0,x_1) =& \frac{1-2t}{t(1-t)}(x-(tx_1+(1-t)x_0)) + (x_1-x_0) \notag \\
  \nabla \log p_t(x|x_0,x_1) =& \frac{tx_1+(1-t)x_0-x}{\sigma^2t(1-t)},
\end{align}
when $g(t)=\sigma$ is a constant, and thus the loss function is written as:
\begin{equation}
  \label{eq:loss-conditional}
 \mathcal{L}_{\text{conditional}}= \mathbb{E}_{t\sim\mathcal{U}(0,1),(x_0,x_1)\sim \pi,x\sim p_t(x|x_0,x_1)}\!\left[\left\|v_\theta(x,t)\!-\!u_t^\circ(x|x_0,x_1)\right\|^2\!+\!\lambda(t)^2 \left\|s_\theta(x,t)\!-\!\nabla \log p_t(x|x_0,x_1)\right\|^2\right].
\end{equation}

\section{Our Method FlowGRN}
\label{sec:methods}

\begin{figure*}[t]
  \centering
  \includegraphics[width=\textwidth]{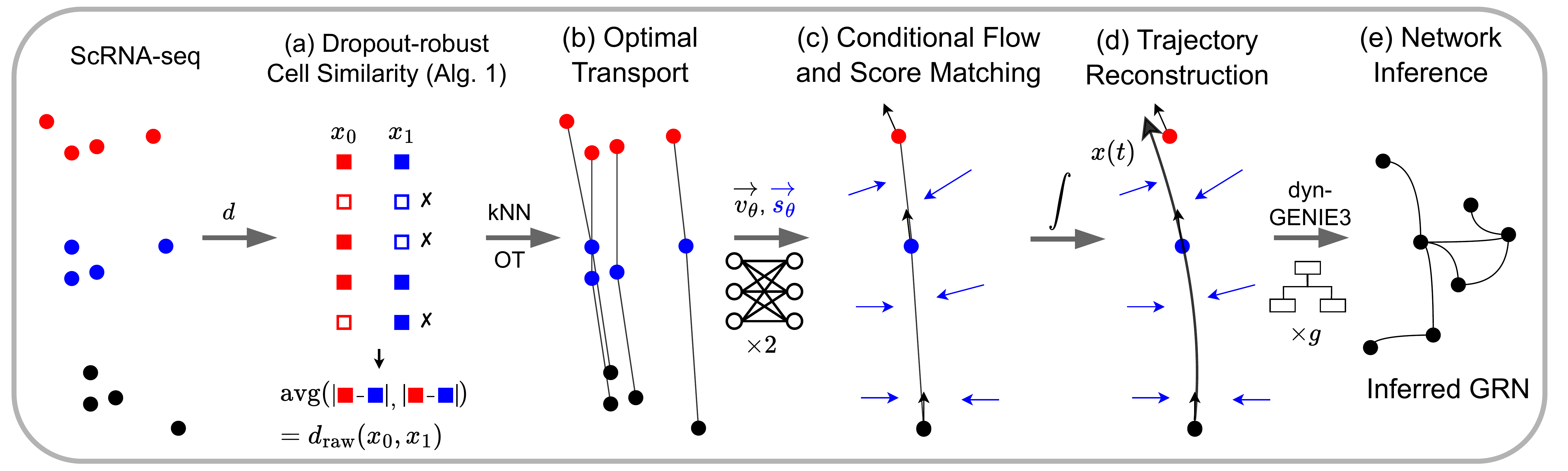}
  \caption{The overall pipeline of FlowGRN.
           (a) Compute $d_{\text{raw}}$ after dropping zeros in each cell pair.
           (b-c) Train CFM $v_\theta$ and score matching $s_\theta$ models to learn the cell dynamics.
           (d) Reconstruct cell trajectories by integrating the learned dynamics.
           (e) Feed trajectories into dynGENIE3 model for GRN inference.
  }
  \Description{The overall pipeline of FlowGRN.
               (a) Compute cell similarity after dropping zeros in each cell pair.
               (b-c) Train CFM and score matching models to learn the cell dynamics.
               (d) Reconstruct cell trajectories by integrating the learned dynamics.
               (e) Feed trajectories into dynGENIE3 model for GRN inference.
               }
  \label{fig:model}
\end{figure*}

To address the identified challenges in GRN inference, we propose FlowGRN, a novel method comprising three parts:
\begin{itemize}
  \item A dropout-robust cell similarity measure that quantifies the geodesic distance between cells in high-dimensional gene space to solve the dropout noise (detailed in Alg.~\ref{alg:cell-sim}).
  \item A conditional flow matching and score matching model that learns the underlying vector fields and reconstructs the cellular trajectories.
  \item The dynGENIE3 model that infers an identifiable GRN from the learned trajectories.
\end{itemize}
Figure~\ref{fig:model} illustrates the overall pipeline of FlowGRN, and Algs.~\ref{alg:flowgrn1}--\ref{alg:flowgrn2} summarize the trajectory reconstruction and GRN inference processes.

\subsection{Dropout-robust cell similarity measure}
\label{ssec:dropout-robust-cell-similarity}

The scalability of [SF]\textsuperscript{2}M comes from the use of the optimal transport (OT) plan $\pi$ to guide the flow.
OT plan $\pi$ is a joint distribution with marginal distributions $p_0,p_1$ that minimizes the cost function $\mathbb{E}_{(x_0,x_1)\sim\pi}[c(x_0,x_1)]$.
The cost function is usually defined as the Euclidean distance $\|x_0-x_1\|_2$, but this metric is severely affected by the dropout noise in the scRNA-seq data and ignores the underlying geometry of the gene expression data manifold.

In Section~\ref{ssec:dropout}, we discuss that only biological zeros and nonzero reads are informative for GRN inference, while technical zeros and sampling zeros are recognized as dropout.
Thus, the zeros are corrupted by dropout, while the non-zeros do not.
This enlightens us to only use non-zeros to quantify the cell similarity.

Given two cells $x,y\in\mathbb{R}^g$, let the set of non-zero genes in $x$ and $y$ be $S_x=\{i|x_i\neq 0\}$ and $S_y=\{i|y_i\neq 0\}$, respectively.
The size of the intersection $S_{x,y}=S_x\cap S_y$ is $|S_{x,y}|$.
We define the dropout-robust cell similarity measure $d_{\rm raw}(x,y)$ as:
\begin{equation}
  d_{\rm raw}(x,y) = \frac{1}{|S_{x,y}|} \sum_{i\in S_{x,y}} |x_i-y_i|.
\end{equation}
Here, the L1 norm is used to avoid distance inflation due to the different sizes of the intersection sets $|S_{x,y}|$.
If $S_{x,y}$ is an empty set, a large $d_{\rm raw}$ will be assigned to indicate dissimilarity.
Although this measure discards the information from the biological zeros, it ensures that the distance is not affected by any dropout noise.

Addressing the geometric structure of the gene expression manifold is not a new idea and has been used in various tasks such as visualization~\cite{mcinnes2018umap,moon2019visualizing}, imputation~\cite{wagner2017knnsmoothing,van2018magic}, and reconstruction of developmental trajectory~\cite{setty2016wishbone}.
We first build a mesh graph $G_{knn}$ from the gene expression matrix $X$ using k-nearest neighbors (kNN) with our cell similarity measure $d_{\rm raw}(x,y)$.
If $G_{knn}$ is not connected, we connect the disconnected components by adding edges between the closest cells in different components.
Then, the final cell similarity $d_{knn}$ is defined as the shortest path distance on the mesh graph $G_{knn}$.
The shortest path distance can be efficiently calculated using Dijkstra's algorithm, or approximated by diffusion~\cite{van2018magic} or random walk~\cite{haghverdi2016diffusion}.
This design brings geometry awareness to the cell similarity measure, and it is robust to dropout noise.
Our algorithm is summarized in Alg.~\ref{alg:cell-sim}, and its effectiveness is tested in Section~\ref{ssec:ablation-cell-similarity-measure}.

\begin{algorithm}[t]
\caption{ComputeCellSimilarity: Dropout-Robust Cell Similarity Measure}\label{alg:cell-sim}
\KwIn{Gene expression matrix $X$, kNN nearest neighbors $k$}
\KwOut{Pairwise distance matrix $d_{knn}$}
\ForEach{cells $x,y\in X$}{
  $S_x\gets\{i\mid x_i\neq0\}$;\  $S_y\gets\{i\mid y_i\neq0\}$;\  $S_{x,y}\gets S_x\cap S_y$\;
  \eIf{$|S_{x,y}|>0$}{
    $d_{\rm raw}(x,y)\gets\frac{1}{|S_{x,y}|}\sum_{i\in S_{x,y}}|x_i-y_i|$\;
  }{
    $d_{\rm raw}(x,y)\gets 10^6$\;
  }
}
Build kNN graph $G_{knn}$ on $X$ with edge weights $d_{\rm raw}$\;
\If{$G_{knn}$ not connected}{
  Connect components by adding edges with minimal $d_{\mathrm{raw}}$\;
}
\ForEach{cells $x,y\in X$}{
  $d_{knn}(x,y)\gets$ DijkstraShortestPath$(G_{knn},x,y)$\;
}
\Return{$d_{knn}$}
\end{algorithm}

\subsection{Trajectory reconstruction with CFM}
\label{ssec:trajectory-reconstruction}
With the cell similarity $d_{knn}$, we can apply the CFM model to learn the underlying vector field $v_\theta(x,t)$ and reconstruct the trajectories.
We first apply Slingshot~\cite{street2018slingshot} to the dataset $X$ and obtain the cell lineages.
Trajectories are reconstructed per lineage to encode the lineage information and maximize parallelism.

Since gene expression values are non-negative, we must ensure that the reconstructed trajectories are also non-negative.
Constraining the boundary of the vector field or enforcing a strictly positive gradient at the boundary are not ideal solutions, as they bring additional complexity and numerical instability.
We solve this issue by converting the expression value $x\in \mathbb{R}^g_{\geq 0}$ to a log domain $x'\in \mathbb{R}^g$ through the following transformation $\mathcal{T}$:
\begin{equation}
  \label{eq:log-like-transformation}
  x' = \mathcal{T}(x) = 
    \begin{cases}
      \log(x+\epsilon) + 1, & \text{if } x+\epsilon < 1 \\
      x+\epsilon, & \text{if } x+\epsilon \geq 1,
    \end{cases}
\end{equation}
where $\epsilon > 0$ is a small positive value to avoid undefined logarithm.
This ensured that the transformation is $C^1$ differentiable at $x=1-\epsilon$ so that training did not get stuck at the discontinuity during optimization.
In addition, the transformation magnifies low expression values and avoids compressing high expression values, thereby capturing cell dynamics at different scales.

We train the CFM and score matching models following the procedure in Section~\ref{ssec:CFM} in the log domain with the loss function $\mathcal{L}_{\text{conditional}}$, conditioned on the OT plan $\pi$ with the cell similarity measure $d_{knn}$ as the cost function.
The trajectories can be recomputed by solving an initial value problem (IVP) starting from $x'(t)$:
\begin{equation}
  x'(t+\Delta t) = x'(t) + \int_t^{t+\Delta t} v_\theta(x'(s),s)\ ds, \quad t_0 \leq t+\Delta t \leq t_{n-1}.
\end{equation}
For each cell $x$ observed at time $t_i$, its transformed state $x'(t_i)=\mathcal{T}(x(t_i))$ serves as an anchor point.
We then generate its trajectory segments by solving the IVP $dx'/dt = v_\theta(x',t)$ forward from $t_i$ to $t_{n-1}$ and backward from $t_i$ to $t_0$.
The workflow of trajectory reconstruction is outlined in Alg.~\ref{alg:flowgrn1}.

\begin{algorithm}[t]
\caption{FlowGRN (Part 1): Trajectory Reconstruction.}
\label{alg:flowgrn1}
\KwIn{Series of snapshots $X=\{X^{(t_0)},\dots,X^{(t_{n-1})}\}$, branch labels $b(x)$, batch size $m$, kNN nearest neighbors $k$}
\KwOut{Reconstructed trajectories $T$, flow network $v_\theta$}
\tcp{Compute cell similarity (see Alg.~\ref{alg:cell-sim})}
$d_{knn}\gets\text{ComputeCellSimilarity}(X, k)$\;

\ForEach{branch $\beta$}{
  \tcp{Modeling the dynamics of branch $\beta$}
  Initialize networks $v_\theta,s_\theta$\;
  \For{$\text{epoch}=1$ \KwTo $E$}{
    \For{$i=0$ \KwTo $n-2$}{
      $X_i\gets X^{(t_i)}\cap\{b(x)=\beta\}$\;
      $X_{i+1}\gets X^{(t_{i+1})}\cap\{b(x)=\beta\}$\;
      Sample batches $\{x_0\}\sim X_i,\;\{x_1\}\sim X_{i+1}$ of size $m$\;
      Compute OT plan $\pi$ between $\{x_0\}$ and $\{x_1\}$ w.r.t.\ cost $d_{knn}$\;
      \ForEach{$(x_0,x_1)\sim\pi$}{
        Sample $t\sim\mathcal U(0,1)$ and $\xi\sim\mathcal N(0,I_g)$\;
        $x_0'= \mathcal{T}(x_0), x_1'=\mathcal{T}(x_1)$ via Eq.~(\ref{eq:log-like-transformation})\;
         $x'\gets t\,x_1'+(1-t)\,x_0'+\sigma\sqrt{t(1-t)}\,\xi$\;
        Compute $u_t^\circ(x'|x_0',x_1'),\;\nabla\log p_t(x'|x_0',x_1')$ via Eq.~(\ref{eq:CFM})\;
      }
    }
    Update $(v_\theta,s_\theta)$ with the loss $\mathcal{L}_{\text{conditional}}$ in Eq.~(\ref{eq:loss-conditional})\;
  }
  \tcp{Reconstructing trajectories}
  \ForEach{cell $x$ where $b(x)=\beta$}{
    Solve $\frac{d x'}{dt}=v_\theta(x',t),\;x'(t_i)=\mathcal{T}(x)$ from $t_0$ to $t_{n-1}$\;
    $x(t)\gets\mathcal{T}^{-1}(x'(t))$\;
    Store $x(t)$
  }
}
Combine all $\{x(t)\}$ over branches into trajectories $T$\;

\Return{$T,\;v_\theta$}
\end{algorithm}

\subsection{GRN inference from the learned trajectory}
\label{ssec:dynGENIE3}
The learned trajectory $x(t)=\mathcal{T}^{-1}(x'(t))$ is a time series of gene expression to be fed into dynGENIE3 to infer the GRN.

dynGENIE3~\cite{huynh2018dyngenie3} is an extension of GENIE3~\cite{huynh2010genie3} that incorporates temporal information into the GRN inference process.
Same as GENIE3, dynGENIE3 uses random forest $f_j$ to model each target gene $j$ in a cell trajectory $x(t)$:
\begin{equation}
  \label{eq:dynGENIE3}
  \frac{dx_j(t)}{dt} = f_j(x(t)) - \alpha_j x_j(t),
\end{equation}
where $\alpha_j$ is the decay rate of gene $j$ and $\frac{dx_j(t)}{dt}$ is approximated by finite differences:
\begin{equation}
  \frac{dx_j(t)}{dt} = \frac{x_j(t+\Delta t) - x_j(t)}{\Delta t}.
\end{equation}
The importance score $\mathcal{I}_j(i)$ of each random forest model $f_j$ is used to quantify the regulatory strength of the gene $i$ on $j$.

To determine the type of regulation, we use the learned vector field $v_\theta$.
Denote $G\in\mathbb{R}^{g\times g\times c}$ as the stack of Jacobians evaluated at the observed cellular states $(x',t)$, such that $G_{i,j,\cdot}$ collects $(\nabla_{x'} v_\theta(x',t))_{i,j}$ across cells.
The regulation type $Gs\in [0,1]^{g\times g}$ is assigned based on a majority vote of the signs in $G$, and the signed adjacency matrix $A$ is defined as:
\begin{equation}
  Gs_{i,j} = \operatorname{sgn}\left(\operatorname{mean}(\operatorname{sgn}\left(G_{i,j,\cdot})\right)\right), \quad A_{i,j}=Gs_{i,j} \cdot \mathcal{I}_j(i).
\end{equation}
Here, the Jacobian is only used to assign the sign of an edge whose strength is determined by dynGENIE3, rather than to determine the existence of the edge.
Also, the asymmetric importance scores $\mathcal{I}_j(i)$ and $\mathcal{I}_i(j)$ yield a directed graph, as the strengths in opposite directions are not generally equal.
This approach allows us to assign regulation types (activation/inhibition) for inferred GRNs, a feature not available from the standard dynGENIE3 output.
The workflow of the GRN inference algorithm is outlined in Alg.~\ref{alg:flowgrn2}.

\begin{algorithm}[t]
\caption{FlowGRN (Part 2): GRN Inference.}
\label{alg:flowgrn2}
\KwIn{Reconstructed trajectories $T$}
\KwOut{Signed adjacency matrix $A\in\mathbb R^{g\times g}$}
\tcp{GRN inference with dynGENIE3}
\For{$j=1$ \KwTo $g$}{
  Fit random forest $f_j$ on $T$ to model 
    $\frac{d x_j(t)}{dt} \approx f_j(x(t))-\alpha_j x_j(t)$ in Eq.~(\ref{eq:dynGENIE3})\;
  Compute importance score $\mathcal I_j(i)$ of gene $i$ on $j$ from $f_j$\;
}
\tcp{Assign edge types using the lineage-specific flow networks}
Initialize $G$ as an empty gradient tensor and $Gs$ as a zero matrix\;
\ForEach{transformed cell $x'$}{
    Append the Jacobian $\nabla_{x'} v_\theta(x',t)$ to $G$\;
}
\For{$j=1$ \KwTo $g$}{
    \For{$i=1$ \KwTo $g$}{
        $Gs_{i,j} \gets \operatorname{sgn}\left(\operatorname{mean}(\operatorname{sgn}\left(G_{i,j,\cdot})\right)\right)$\;
        $A_{i,j} \gets Gs_{i,j} \cdot \mathcal I_j(i)$\;
    }
}
\Return{$A$}
\end{algorithm}

\subsection{Model implementation}
\label{ssec:implementation}
A pictorial overview of the FlowGRN model is shown in Figure~\ref{fig:model}, and the detailed algorithms of our new similarity measure and FlowGRN pipeline are outlined in Algs.~\ref{alg:cell-sim}--\ref{alg:flowgrn2}.
We first compute the pairwise cell similarity $d_{knn}$ using Alg.~\ref{alg:cell-sim} on the gene expression matrix $X$.
Then, for each lineage $\beta$, we initialize the flow matching $v_\theta$ and score matching model $s_\theta$.
For each successive pair of snapshots $X^{(t_i)}, X^{(t_{i+1})}$, we compute the OT plan $\pi$ between two snapshots using the cell similarity $d_{knn}$, and then sample $(x_0,x_1)\sim\pi$.
Samples are transformed to the log domain via $\mathcal{T}$ in Eq.~(\ref{eq:log-like-transformation}) and train the neural ODEs following Eqs.~(\ref{eq:CFM}) and~(\ref{eq:loss-conditional}).
We reconstruct the trajectories by solving the IVP for each cell and inverse transform the trajectories back to the original gene expression space.
Finally, we combine all trajectories $T$ from different lineages and feed them into dynGENIE3 to infer the signed GRN $A$.

\section{Experiments}
\label{sec:experiments}

\subsection{Datasets and evaluation metrics}
\label{ssec:datasets}

We tested our method on the BEELINE~\cite{pratapa2020benchmarking} benchmark, which includes 6 synthetic datasets, 4 curated datasets, and 7 experimental datasets.
Synthetic data sets include linear (LI), linear long (LL), cyclic (CY), bifurcating (BF), bifurcating converging (BFC), and trifurcating (TF), capturing common dynamics such as long cascades and furcating branches.
Curated datasets include mCAD, VSC, HSC, and GSD, which mimic the reported biological processes.
Both types of datasets are simulated with BoolODE~\cite{pratapa2020benchmarking} to generate 10 sets of simulated scRNA-seq data with 2,000 cells.
Their network statistics are shown in Table~\ref{tab:syn-cur-network-statistics} in the Appendix.

Experimental datasets are real gene expression profiles from sequencing data.
There are 7 sets of gene expression profiles from human and mouse cells, namely hESC, hHep, mDC, mESC, mHSC-E, mHSC-GM, and mHSC-L.
The top 500 (1,000) high-variance genes and their corresponding transcription factors are selected from each dataset, matching the ``TFs + 500 (1,000) genes'' in the BEELINE benchmark.
Although BEELINE provides cell-type-specific and non-cell-type-specific GRNs as reference networks, these reference networks have not been updated since 2020.
We retrieved the gene regulations from the DoRothEA~\cite{garcia2019dorothea} and CollecTRI~\cite{muller2023collectri} databases and used them as reference networks.
Their data statistics are shown in Table~\ref{tab:experimental-network-statistics}.

\begin{table}[!t]
\centering
\caption{Network Statistics of the Experimental Datasets}
\label{tab:experimental-network-statistics}
\begin{tabular}{lrrrr}
\toprule[1pt]\midrule[0.3pt]
Dataset      & \# Cells & \# Genes & \# Edges & \begin{tabular}[c]{@{}@{}r@{}}Network\\Density\\ ($\times 10^{-3}$)\end{tabular} \\
\hline
\multicolumn{5}{l}{\textbf{TFs + 500 genes}} \\
\quad hESC     & 758      & 1283     & 20007    & 12.15                                                                      \\
\quad hHep     & 425      & 1209     & 16915    & 11.57                                                                      \\
\quad mDC      & 383      & 1113     & 2854     & 2.30                                                                       \\
\quad mESC     & 421      & 501      & 139      & 0.55                                                                       \\
\quad mHSC-E   & 1071     & 763      & 945      & 1.62                                                                       \\
\quad mHSC-GM  & 889      & 698      & 676      & 1.39                                                                       \\
\quad mHSC-L   & 847      & 624      & 442      & 1.14                                                                       \\
\hline
\multicolumn{5}{l}{\textbf{TFs + 1,000 genes}} \\
\quad hESC    & 758      & 1783     & 26553    & 8.35                                                                       \\
\quad hHep    & 425      & 1709     & 22007    & 7.53                                                                       \\
\quad mDC     & 383      & 1613     & 3404     & 1.31                                                                       \\
\quad mESC    & 421      & 1001     & 416      & 0.42                                                                       \\
\quad mHSC-E  & 1071     & 1263     & 1135     & 0.71                                                                       \\
\quad mHSC-GM & 889      & 1198     & 895      & 0.62                                                                       \\
\quad mHSC-L  & 847      & 1124     & 729      & 0.58                                                                       \\
\midrule[0.3pt]\bottomrule[1pt]
\end{tabular}
\end{table}

We compare FlowGRN with the baselines on the BEELINE benchmark, which includes LEAP~\cite{specht2017leap}, SCODE~\cite{matsumoto2017scode}, GRISLI~\cite{aubin2020grisli}, GRNVBEM~\cite{sanchez2018grnvbem}, SINCERITIES~\cite{papili2018sincerities}, Scribe~\cite{qiu2020scribe}, GENIE3~\cite{huynh2010genie3}, and GRNBOOST2~\cite{moerman2019grnboost2}.
Their descriptions are listed in Table~\ref{tab:models_compare} in Appendix~\ref{sec:appendix}.
Among the listed baselines, only SCODE, GRNVBEM, SINCERITIES, and FlowGRN support inference of edge types (activation/inhibition).
Here, we excluded SCNS~\cite{woodhouse2018scns} as it requires prior knowledge of GRN.
Furthermore, we excluded ppcor~\cite{kim2015ppcor} and PIDC~\cite{chan2017pidc} because they cannot determine the direction of regulations.

We evaluate the performance of FlowGRN and the baselines on the BEELINE benchmark using two metrics: Area under Precision-Recall Curve (AUPRC) and Early Precision Ratio  (EPR).
For a model output $A$ and a threshold $\tau$, the numbers of true positive $TP(\tau)$, false positive $FP(\tau)$, true negative $TN(\tau)$, and false negative $FN(\tau)$ edges are counted.
Precision is defined as $P(\tau)=\frac{TP(\tau)}{TP(\tau)+FP(\tau)}$, and recall is defined as $R(\tau)=\frac{TP(\tau)}{TP(\tau)+FN(\tau)}$.
The AUPRC is calculated by integrating the precision over all thresholds $\tau$:
\begin{equation}
  \text{AUPRC} = \int_0^1 P(\tau) dR(\tau).
\end{equation}
The EPR is defined as the precision among the top $k$ edges, and the definition of precision is the same as above.
Both metrics emphasize the precision of the inferred GRN and minimize the false positive rate, with higher values indicating better performance.
However, EPR is sensitive to the number of edges in the ground truth network, especially when the network is sparse.

\subsection{Experimental Setup}
\label{ssec:experimental-setup}
We first construct a series of snapshots for each dataset by clustering cells and ordering each cell cluster by their average pseudotime.
Details of the preprocessing are described in Appendix~\ref{ssec:appendix-preprocessing}.
During trajectory reconstruction, $k=15$ for the kNN graph $G_{knn}$, $\sigma$ is set to 0.1 for the Brownian motion diffusion, $\lambda(t)$ is set to 1 for the score matching loss, and $\epsilon$ is set to $10^{-6}$ for the log-like transformation.
Exact OT plans are computed using the Python Optimal Transport (POT) library~\cite{flamary2021pot}.
During trajectory reconstruction, $v_\theta, s_\theta$ in the [SF]\textsuperscript{2}M model are implemented as MLPs with 5 hidden layers, each with 128 hidden units and SELU activations.
Both neural networks are optimized by AdamW with a learning rate of $10^{-4}$ decayed by a factor of 2 every 100 epochs.
Models are trained for 1,000 epochs with a batch size of 64 samples.
Trajectories are integrated using a DOPRI5 neural ODE solver~\cite{poli2020torchdyn} with an absolute and relative tolerance of $10^{-7}$, sampled at 5 time points between each pair of snapshots.
All neural network models are trained on a single NVIDIA V100 SXM2 GPU with 16GB memory, although the peak memory usage is less than 1GB and can be run on a domestic computer.

In GRN inference, dynGENIE3 is used with the default parameters: 1,000 random forest trees, each trained with $\sqrt{g}$ genes.
dynGENIE3 is trained on a computing node with 128 CPUs at 2.6GHz and 256GB memory.
All baselines are also run on the same platform within a given 48-hour time interval, using their default parameters wrapped in a Singularity container to ensure reproducibility.
Each method, including FlowGRN, is run 10 times with different random seeds on each dataset, and the average and standard deviation of AUPRC and EPR are reported.
Model elapsed times are recorded and reported in Table~\ref{tab:time} for scalability analysis.

\section{Results and Discussion}
\label{sec:results-discussion}

\subsection{FlowGRN infers identifiable GRNs in simulated data}
\label{ssec:results-directed}

\begin{figure}[!t]
  \centering
  \includegraphics[width=0.8\linewidth]{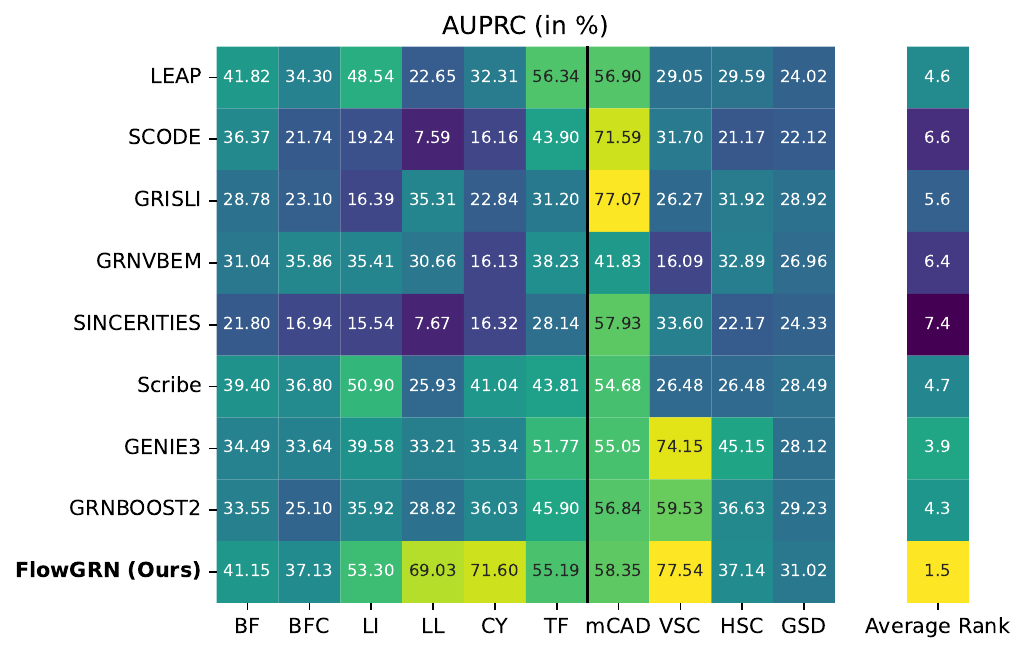}
  \Description{AUPRC and average rank of directed GRN inference models on different simulated datasets.}
  \caption{Average AUPRC and its average rank of directed GRN inference models on different simulated datasets out of 10 runs.
           Synthetic and curated datasets are separated by a vertical line.}
  \label{fig:directed}
\end{figure}

\begin{figure}[!t]
  \centering
  \includegraphics[width=0.8\linewidth]{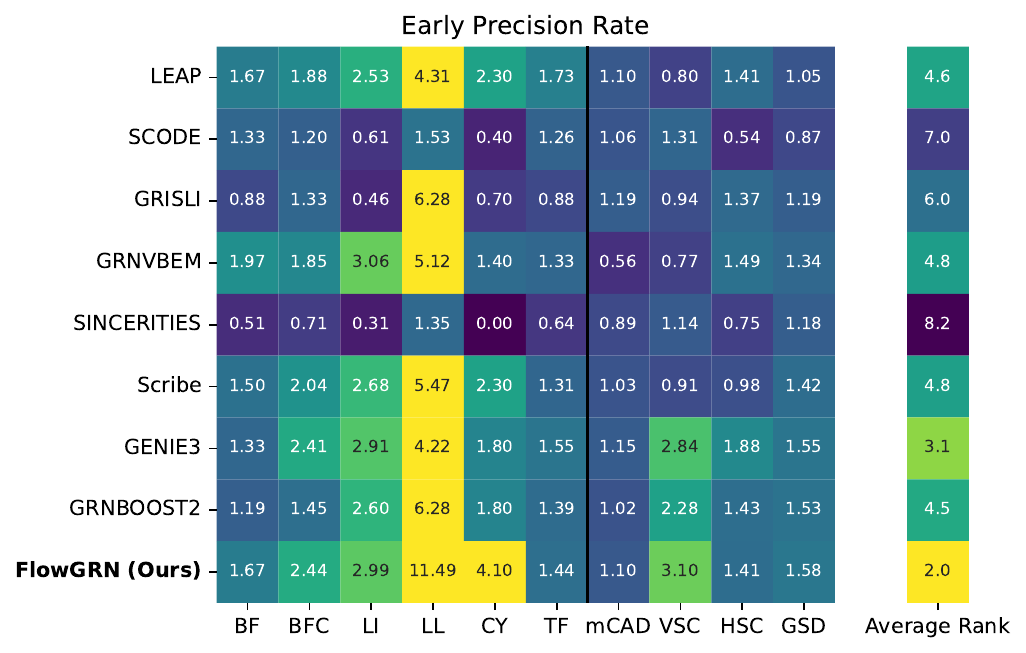}
  \Description{EPR and average rank of directed GRN inference models on different simulated datasets.}
  \caption{Average EPR and its average rank of directed GRN inference models on different simulated datasets out of 10 runs.
           Synthetic and curated datasets are separated by a vertical line.
           Color scale is capped at 4 for better visualization.
           }
  \label{fig:directed-epr}
\end{figure}

We first evaluate FlowGRN on the BEELINE benchmark using both synthetic and curated datasets, and the average AUPRC and EPR are shown in Figures~\ref{fig:directed} and \ref{fig:directed-epr}, respectively.
The standard deviations are shown in Figures~\ref{fig:directed-std} and \ref{fig:directed-epr-std} in the Appendix.
FlowGRN has the highest average rank of AUPRC (1.5) and EPR (2.0) among all models, significantly surpassing the second-best model GENIE3 with average ranks of 3.9 in AUPRC and 3.1 in EPR.

Although FlowGRN adopts dynGENIE3, a temporal extension of GENIE3,
it outperforms GENIE3, especially in the LL and CY datasets.
The LL dataset features long cascades, indicating that FlowGRN effectively captures long-range gene expression dependencies.
The CY dataset is characterized by cyclic dynamics, thus determining the edge direction is crucial for the GRN inference.
Incorporating temporal information enables FlowGRN to capture the principal direction of cyclic dynamics, which static models like GENIE3 struggle with.

\subsection{FlowGRN determines edge types in simulated data}
\label{ssec:results-signed}

\begin{figure}[!t]
  \centering
  \includegraphics[width=0.8\linewidth]{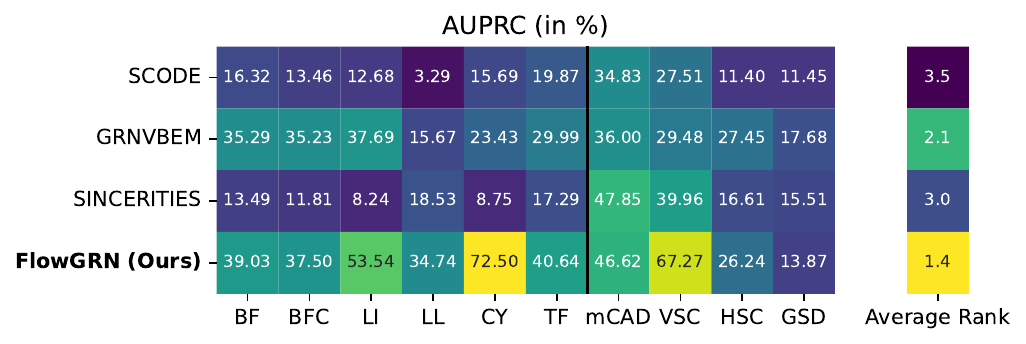}
  \Description{AUPRC of directed GRN inference models with edge types on different simulated datasets.}
  \caption{Average AUPRC of directed GRN inference models with edge types on different simulated datasets out of 10 runs.
           Synthetic and curated datasets are separated by a vertical line.
           }
  \label{fig:signed}
\end{figure}

We further test FlowGRN's ability to determine the activation and inhibition edges on top of the directed GRN inference.
Figure~\ref{fig:signed} shows the average AUPRC of FlowGRN and the baselines on the BEELINE benchmark with edge types, and their standard deviations are reported in Figure~\ref{fig:signed-std} in the Appendix.
EPR is not applicable here as GRN inference with edge type is a multi-class classification problem.
Note that there are fewer baselines capable of inferring edge types.
FlowGRN achieves the best performance in 7 out of 10 datasets, and its performance remarkably surpasses the second-best model, GRNVBEM.

\subsection{FlowGRN is a top-tier model on experimental data}
\label{ssec:results-experimental}

\begin{table*}[t]
\centering
\caption{AUPRC ($\times 10^{-3}$) of directed GRN inference models on different experimental datasets out of 10 runs.
         The \textbf{\underline{best}} and \underline{second-best} results are highlighted in bold and underlined, respectively.
         ``OT'' and ``NE'' mean ``over time in a 48-hour time span'' and ``numerical error'', respectively.
         }
\label{tab:AUPR-real}
\begin{tabular}{lrrrrrrr}
\toprule[1pt]\midrule[0.3pt]
dataset & hESC & hHep & mDC & mESC & mHSC-E & mHSC-GM & mHSC-L \\
\midrule
\multicolumn{8}{l}{\textbf{TFs + 500 genes}} \\
 		\quad LEAP 		    & 11.91$\pm$0.00 		& \textbf{\underline{12.99$\pm$0.00}} 		& 2.38$\pm$0.00 		& \textbf{\underline{5.36$\pm$0.00}} 		& 1.94$\pm$0.00 		& 2.00$\pm$0.00 		& 1.28$\pm$0.00 \\
 		\quad SCODE 		  & 10.00$\pm$0.10 		& 10.56$\pm$0.04 		& 2.01$\pm$0.04 		& 0.84$\pm$0.03 		& 1.50$\pm$0.02 		& 1.30$\pm$0.03 		& 0.94$\pm$0.01 \\
 		\quad GRISLI 		  & 11.05$\pm$0.00 		& 10.57$\pm$0.00 		& \textbf{\underline{2.95$\pm$0.00}} 		& 0.51$\pm$0.00 		& 1.65$\pm$0.00 		& 1.21$\pm$0.00 		& 1.29$\pm$0.00 \\
    \quad GRNVBEM 		& OT             		& OT             		& 2.34$\pm$0.00 		& 0.55$\pm$0.00 		& 1.62$\pm$0.00 		& 1.39$\pm$0.00 		& 1.14$\pm$0.00 \\
 		\quad SINCERITIES & \underline{12.47$\pm$0.00} 		& 10.60$\pm$0.00 		& \underline{2.69$\pm$0.00} 		& 0.51$\pm$0.00 		& 1.49$\pm$0.00 		& 1.37$\pm$0.00 		& 1.18$\pm$0.00 \\
 		\quad Scribe 		  & 10.67$\pm$0.00 		& 9.64$\pm$0.00 		& 1.83$\pm$0.00 		& 0.53$\pm$0.00 		& 1.65$\pm$0.00 		& 1.38$\pm$0.00 		& 1.12$\pm$0.00 \\
		\quad GENIE3 		  & 10.50$\pm$0.01 		& 9.75$\pm$0.02 		& 2.10$\pm$0.01 		& \underline{3.01$\pm$0.10} 		& \underline{3.02$\pm$0.43} 		& \underline{2.44$\pm$0.08} 		& \underline{1.95$\pm$0.10} \\
 		\quad GRNBOOST2 	& 11.97$\pm$0.02 		& 11.42$\pm$0.02 		& 2.32$\pm$0.01 		& 2.45$\pm$0.62 		& 1.78$\pm$0.04 		& 1.93$\pm$0.11 		& 1.63$\pm$0.37 \\
 		\quad FlowGRN (Ours)     & \textbf{\underline{12.53$\pm$0.11}} 		& \underline{12.66$\pm$0.19} 		& 2.32$\pm$0.03 		& 1.83$\pm$0.15 		& \textbf{\underline{3.36$\pm$0.23}} 		& \textbf{\underline{3.33$\pm$0.14}} 		& \textbf{\underline{2.35$\pm$0.14}} \\
\midrule
\multicolumn{8}{l}{\textbf{TFs + 1,000 genes}} \\
   	\quad LEAP 		    & \underline{8.79$\pm$0.00} 		& \textbf{\underline{8.99$\pm$0.00}} 		& 1.38$\pm$0.00 		& \textbf{\underline{1.89$\pm$0.00}} 		& 1.01$\pm$0.00 		& 1.01$\pm$0.00 		& 0.66$\pm$0.00 \\
 		\quad SCODE 		  & 6.68$\pm$0.08 		& 6.74$\pm$0.08 		& 1.08$\pm$0.01 		& 0.52$\pm$0.01 		& 0.63$\pm$0.01 		& 0.58$\pm$0.01 		& 0.47$\pm$0.01 \\
 		\quad GRISLI 		  & 7.71$\pm$0.00 		& 7.01$\pm$0.00 		& \textbf{\underline{1.72$\pm$0.00}} 		& 0.42$\pm$0.00 		& 0.72$\pm$0.00 		& 0.55$\pm$0.00 		& 0.63$\pm$0.00 \\
    \quad GRNVBEM 		& OT             		& OT             		& OT             		& 0.42$\pm$0.00 		& OT             		& OT             		& 0.58$\pm$0.00 \\
 		\quad SINCERITIES & 8.57$\pm$0.00 		& 6.95$\pm$0.00 		& \underline{1.47$\pm$0.00} 		& 0.37$\pm$0.00 		& NE 		& 0.62$\pm$0.00 		& NE \\
 		\quad Scribe 		  & 7.30$\pm$0.00 		& 6.17$\pm$0.00 		& 0.99$\pm$0.00 		& 0.48$\pm$0.00 		& 0.71$\pm$0.00 		& 0.65$\pm$0.00 		& 0.63$\pm$0.00 \\
		\quad GENIE3 		  & 7.19$\pm$0.01 		& 6.25$\pm$0.02 		& 1.17$\pm$0.01 		& \underline{1.11$\pm$0.03} 		& \underline{1.30$\pm$0.08} 		& \underline{1.14$\pm$0.04} 		& \underline{0.88$\pm$0.02} \\
 		\quad GRNBOOST2   & 8.29$\pm$0.01 		& 7.48$\pm$0.02 		& 1.33$\pm$0.01 		& 0.88$\pm$0.07 		& 0.82$\pm$0.02 		& 0.88$\pm$0.05 		& 0.70$\pm$0.05 \\
 		\quad FlowGRN (Ours)	    & \textbf{\underline{8.82$\pm$0.15}} 		& \underline{8.98$\pm$0.76} 		& 1.27$\pm$0.02 		& 0.81$\pm$0.05 		& \textbf{\underline{1.60$\pm$0.11}} 		& \textbf{\underline{1.56$\pm$0.07}} 		& \textbf{\underline{0.93$\pm$0.04}} \\
\midrule[0.3pt]\bottomrule[1pt]
\end{tabular}
\end{table*}

\begin{table*}[t]
\centering
\caption{EPR of directed GRN inference models on different experimental datasets out of 10 runs.
         The \textbf{\underline{best}} and \underline{second-best} results are highlighted in bold and underlined, respectively.
         ``OT'' and ``NE'' mean ``over time in a 48-hour time span'' and ``numerical error'', respectively.}
\label{tab:EPR-real}
\begin{tabular}{lrrrrrrr}
\toprule[1pt]\midrule[0.3pt]
dataset & hESC & hHep & mDC & mESC & mHSC-E & mHSC-GM & mHSC-L \\
\midrule
\multicolumn{8}{l}{\textbf{TFs + 500 genes}} \\
 		\quad LEAP 		    & 0.94$\pm$0.00 		& \textbf{\underline{1.84$\pm$0.00}} 		& 1.98$\pm$0.00 		& \textbf{\underline{38.97$\pm$0.00}} 		& \textbf{\underline{5.87$\pm$0.00}} 		& 3.20$\pm$0.00 		& 1.99$\pm$0.00 \\
 		\quad SCODE 		  & 0.39$\pm$0.08 		& 0.99$\pm$0.04 		& 0.00$\pm$0.00 		& 0.00$\pm$0.00 		& 0.00$\pm$0.00 		& 2.13$\pm$0.00 		& 0.00$\pm$0.00 \\
 		\quad GRISLI 		  & 0.94$\pm$0.00 		& 0.91$\pm$0.00 		& \textbf{\underline{2.13$\pm$0.00}} 		& 0.00$\pm$0.00 		& 1.96$\pm$0.00 		& 1.07$\pm$0.00 		& 5.98$\pm$0.00 \\
    \quad GRNVBEM 		& OT             		& OT 		            & 0.76$\pm$0.00 		& 0.00$\pm$0.00 		& 1.96$\pm$0.00 		& 1.07$\pm$0.00 		& 0.00$\pm$0.00 \\
 		\quad SINCERITIES & \textbf{\underline{1.34$\pm$0.00}} 		& 0.51$\pm$0.00 		& 1.06$\pm$0.00 		& 0.00$\pm$0.00 		& 0.65$\pm$0.00 		& 0.00$\pm$0.00 		& 0.00$\pm$0.00 \\
 		\quad Scribe 		  & 0.67$\pm$0.00 		& 0.61$\pm$0.00 		& 0.00$\pm$0.00 		& 0.00$\pm$0.00 		& 1.30$\pm$0.00 		& 1.07$\pm$0.00 		& 0.00$\pm$0.00 \\
		\quad GENIE3 		  & 0.82$\pm$0.02 		& 1.07$\pm$0.02 		& \underline{2.05$\pm$0.23} 		& 3.90$\pm$6.28 		& 5.41$\pm$0.44 		& \underline{8.10$\pm$0.55} 		& 5.98$\pm$0.00 \\
 		\quad GRNBOOST2   & 0.78$\pm$0.02 		& 0.95$\pm$0.04 		& 1.43$\pm$0.29 		& \underline{11.69$\pm$12.92} 	& 2.93$\pm$0.98 		& \underline{8.10$\pm$2.02} 		& \textbf{\underline{6.98$\pm$1.69}} \\
 		\quad FlowGRN (Ours)     & \underline{1.09$\pm$0.10} 		& \underline{1.40$\pm$0.11} 		& 0.52$\pm$0.22 		& 0.00$\pm$0.00 		& \underline{5.80$\pm$0.57} 		& \textbf{\underline{9.38$\pm$0.67}} 		& \textbf{\underline{6.98$\pm$1.41}} \\
\midrule
\multicolumn{8}{l}{\textbf{TFs + 1,000 genes}} \\
 		\quad LEAP 		    & 1.28$\pm$0.00 		& \textbf{\underline{2.36$\pm$0.00}} 		& \textbf{\underline{2.92$\pm$0.00}} 		& \textbf{\underline{46.32$\pm$0.00}} 		& 8.67$\pm$0.00 		& 5.38$\pm$0.00 		& 0.00$\pm$0.00 \\
 		\quad SCODE 		  & 0.27$\pm$0.07 		& 1.02$\pm$0.05 		& 0.00$\pm$0.00 		& 0.00$\pm$0.00 		& 0.00$\pm$0.00 		& 4.12$\pm$1.21 		& 0.00$\pm$0.00 \\
 		\quad GRISLI 		  & 0.82$\pm$0.00 		& 0.93$\pm$0.00 		& \textbf{\underline{2.92$\pm$0.00}} 		& 0.00$\pm$0.00 		& 1.24$\pm$0.00 		& 0.00$\pm$0.00 		& 2.38$\pm$0.00 \\
    \quad GRNVBEM 		& OT             		& OT             		& OT             		& 0.00$\pm$0.00 		& OT             		& OT             		& 0.00$\pm$0.00 \\
 		\quad SINCERITIES & \underline{1.42$\pm$0.00 }		& 0.58$\pm$0.00 		& 0.22$\pm$0.00 		& 0.00$\pm$0.00 		& NE 		  & 0.00$\pm$0.00 		& NE \\
 		\quad Scribe 		  & 0.69$\pm$0.00 		& 0.60$\pm$0.00 		& 0.22$\pm$0.00 		& \underline{11.58$\pm$0.00} 		& 2.48$\pm$0.00 		& 0.00$\pm$0.00 		& 0.00$\pm$0.00 \\
		\quad GENIE3 		  & 1.10$\pm$0.02 		& 1.31$\pm$0.04 		& 2.87$\pm$0.25 		& 5.79$\pm$4.73 		& \underline{9.16$\pm$1.20} 		& \underline{13.26$\pm$1.25} 		& 7.13$\pm$0.00 \\
 		\quad GRNBOOST2   & 0.96$\pm$0.04 		& 1.09$\pm$0.06 		& 1.82$\pm$0.40 		& 11.00$\pm$6.93 		& 6.81$\pm$1.46 		& 11.47$\pm$2.42 		& \textbf{\underline{8.56$\pm$1.66}} \\
 		\quad FlowGRN (Ours)	    & \textbf{\underline{1.47$\pm$0.20}} 		& \underline{2.23$\pm$0.53} 		& 0.90$\pm$0.33 		& 0.58$\pm$1.83 		& \textbf{\underline{11.89$\pm$0.64}} 		& \textbf{\underline{15.77$\pm$1.13}} 		& \underline{7.37$\pm$0.75} \\
\midrule[0.3pt]\bottomrule[1pt]
\end{tabular}
\end{table*}

We compared FlowGRN with the baselines on 7 experimental datasets, and their performances are listed in Tables~\ref{tab:AUPR-real}--\ref{tab:EPR-real}.
The main difficulties of these datasets are the high dimensionality and the presence of dropout, as discussed in Sections~\ref{ssec:dimensionality} and \ref{ssec:dropout}.
Most baselines successfully completed the tasks, except for GRNVBEM, which failed to complete half of its tasks within a 48-hour time span, and SINCERITIES, which exhibits numerical errors on the mHSC-E and mHSC-L datasets.

LEAP, GENIE3, and FlowGRN are the top-tier models on both metrics.
FlowGRN ranks first or second in 10 of 14 tasks for both AUPRC and EPR, compared with 5 and 6 tasks for LEAP and 8 and 4 tasks for GENIE3, respectively.
All of them adopt the idea of feature selection in their methodologies (Section~\ref{ssec:dimensionality}), suggesting that feature selection is crucial for GRN inference in high-dimension and sparse interactions.

Our model demonstrates scalability up to 1,783 genes.
However, it achieves only average performance in the mDC and mESC datasets.
This can be attributed to the low number of cells (383 and 421 cells) in both datasets, which is approximately half the number of cells in the other datasets.
Insufficient cell samples may result in inadequate information for the model to accurately capture complex gene interactions, particularly in deep learning models.
We believe that this issue can be addressed by simplifying the model's complexity or incorporating biological priors into the flow matching model.

\subsection{Computational scalability}

The model run times are reported in Table~\ref{tab:time}.
Our model is as scalable as GENIE3 and GRNBOOST2, which can be trained in less than 10 minutes.
However, different methods have various levels of parallelization, and we have optimized the implementation for GENIE3, GRNBOOST2, and FlowGRN on multiple-core computation.

\begin{table}[!t]
\centering
\caption{Average elapsed time (seconds) of different GRN inference models in various dataset sizes}\label{tab:time}
\begin{tabular}{
    >{\raggedleft\arraybackslash}p{0.075\linewidth}%
    >{\raggedleft\arraybackslash}p{0.075\linewidth}%
    >{\raggedleft\arraybackslash}p{0.075\linewidth}%
    >{\raggedleft\arraybackslash}p{0.075\linewidth}%
    >{\raggedleft\arraybackslash}p{0.075\linewidth}%
    >{\raggedleft\arraybackslash}p{0.075\linewidth}%
    >{\raggedleft\arraybackslash}p{0.075\linewidth}%
    >{\raggedleft\arraybackslash}p{0.075\linewidth}%
    >{\raggedleft\arraybackslash}p{0.075\linewidth}%
    >{\raggedleft\arraybackslash}p{0.075\linewidth}%
}
\toprule[1pt]\midrule[0.3pt]
\# genes & LEAP & SCODE & GRISLI & GRN-VBEM & SINCER-ITIES & Scribe & GENIE3 & GRN-BOOST2 & FlowGRN \\
\hline
        501 & 44.20 & 2424.83 & 1660.29 & 280.16 & 121.80 & 1627.46 & 38.34 & 20.15 & 84.10 \\
        624 & 155.84 & 3414.29 & 1930.55 & 303.83 & 206.57 & 6527.50 & 86.63 & 16.94 & 104.80 \\
        698 & 190.48 & 4062.10 & 2846.68 & 414.96 & 299.78 & 8648.80 & 111.44 & 18.81 & 117.17 \\
        763 & 203.75 & 4479.20 & 4082.15 & 555.18 & 304.91 & 13190.50 & 151.49 & 22.66 & 128.26 \\
        1001 & 139.61 & 4420.80 & 5789.50 & 1284.02 & 458.13 & 6490.50 & 82.61 & 32.52 & 167.92 \\
        1113 & 144.60 & 4918.70 & 6317.20 & 1831.28 & 782.26 & 7416.80 & 90.934 & 22.31 & 187.07 \\
        1124 & 253.55 & 5862.20 & 9271.00 & 1862.92 & - & 22202.30 & 204.17 & 37.35 & 188.92 \\
        1198 & 327.18 & 6322.10 & 11711.20 & - & 886.09 & 25344.70 & 236.54 & 40.62 & 201.37 \\
        1209 & 291.89 & 5431.50 & 8433.50 & - & 882.68 & 9994.00 & 91.46 & 25.56 & 203.36 \\
        1263 & 358.72 & 7180.30 & 15125.00 & - & - & 35624.30 & 325.91 & 37.25 & 212.09 \\
        1283 & 363.35 & 6685.20 & 12561.40 & - & 1157.01 & 25166.50 & 191.59 & 32.61 & 215.38 \\
        1613 & 417.06 & 6859.70 & 16224.90 & - & 1313.31 & 15535.30 & 149.80 & 31.65 & 270.81 \\
        1709 & 429.82 & 7256.10 & 19567.90 & - & 1285.11 & 19758.00 & 151.11 & 34.47 & 287.16 \\
        1783 & 465.11 & 9019.50 & 27275.80 & - & 1442.53 & 47802.70 & 308.98 & 44.18 & 299.60 \\
\midrule[0.3pt]\bottomrule[1pt]
\end{tabular}
\end{table}

\subsection{Trajectory reconstruction is irreplaceable in FlowGRN}
\label{ssec:ablation-trajectory}

\begin{table*}[t]
\centering
\small
\caption{AUPRC ($\times 10^{-3}$) of FlowGRN and GENIE3 trained with and without trajectories in different experimental datasets out of 10 runs.
         The \textbf{\underline{best}} and \underline{second-best} results are highlighted in bold and underlined, respectively.
         }
\label{tab:AUPR-ablation-traj}  
\begin{tabular}{lrrrrrrr}
\toprule[1pt]\midrule[0.3pt]
dataset & hESC & hHep & mDC & mESC & mHSC-E & mHSC-GM & mHSC-L \\
\midrule
\multicolumn{8}{l}{\textbf{TFs + 500 genes}} \\
		\; GENIE3             		& 10.497$\pm$0.012 		& 9.754$\pm$0.015 		& 2.097$\pm$0.013 		& \textbf{\underline{3.006$\pm$0.100}} 		& \underline{3.018$\pm$0.429} 		& \underline{2.442$\pm$0.083} 		& \underline{1.950$\pm$0.101} \\
 		\; \; W/ traj. 		& \underline{12.154$\pm$0.000} 		& 11.572$\pm$0.000 		& \underline{2.304$\pm$0.000} 		& 0.554$\pm$0.000 		& 1.623$\pm$0.000 		& 1.388$\pm$0.000 		& 1.135$\pm$0.000 \\
 		\; FlowGRN 		          & \textbf{\underline{12.526$\pm$0.112}} 		& \underline{12.660$\pm$0.186} 		& \textbf{\underline{2.315$\pm$0.030}} 		& \underline{1.826$\pm$0.152} 		& \textbf{\underline{3.364$\pm$0.233}} 		& \textbf{\underline{3.333$\pm$0.145}} 		& \textbf{\underline{2.348$\pm$0.139}} \\
 		\; \; W/o traj. 	& 10.250$\pm$0.190 		& \textbf{\underline{13.360$\pm$0.564}} 		& 2.014$\pm$0.055 		& 1.254$\pm$0.182 		& 1.236$\pm$0.055 		& 1.230$\pm$0.062 		& 0.976$\pm$0.095 \\
\midrule
\multicolumn{8}{l}{\textbf{TFs + 1,000 genes}} \\
		\; GENIE3             		& 7.192$\pm$0.012 		& 6.246$\pm$0.017 		& 1.172$\pm$0.006 		& \textbf{\underline{1.105$\pm$0.033}} 		& \underline{1.296$\pm$0.083} 		& \underline{1.142$\pm$0.037} 		& \underline{0.877$\pm$0.016} \\
 		\; \; W/ traj. 		& \underline{8.352$\pm$0.000} 		& 7.535$\pm$0.000 		& \textbf{\underline{1.308$\pm$0.000}} 		& 0.415$\pm$0.000 		& 0.712$\pm$0.000 		& 0.624$\pm$0.000 		& 0.577$\pm$0.000 \\
 		\; FlowGRN 		          & \textbf{\underline{8.815$\pm$0.147}} 		& \textbf{\underline{8.978$\pm$0.761}} 		& \underline{1.270$\pm$0.024} 		& 0.808$\pm$0.048 		& \textbf{\underline{1.602$\pm$0.109}} 		& \textbf{\underline{1.556$\pm$0.067}} 		& \textbf{\underline{0.931$\pm$0.041}} \\
 		\; \; W/o traj. 	& 7.313$\pm$0.218 		& \underline{7.814$\pm$0.273} 		& 1.044$\pm$0.021 		& \underline{0.962$\pm$0.203} 		& 0.541$\pm$0.012 		& 0.546$\pm$0.030 		& 0.504$\pm$0.026 \\
\midrule[0.3pt]\bottomrule[1pt]
\end{tabular}
\end{table*}

We performed an ablation study on the use of reconstructed trajectories in the GENIE3 and FlowGRN models, and the results are shown in Table~\ref{tab:AUPR-ablation-traj}.
GENIE3 initially relies on static gene expression data, but we may also use the reconstructed trajectories as augmented inputs, as the set of all trajectories contains static gene expression data.
The latter approach is labeled as ``GENIE3 (w/ traj.)''.
In FlowGRN, the trajectories reconstructed by the CFM model serve as input to the dynGENIE3 model.
However, we may also use the trained vector field $v_\theta$ from the CFM model to reconstruct GRN directly from its Jacobian $\nabla_{x'} v_\theta$, labeled as ``FlowGRN (w/o traj.)''.

This ablation study demonstrates that trajectory reconstruction is an indispensable component in the FlowGRN model, as the performance of FlowGRN without trajectories is significantly lower than that with trajectories.
This is because the chain rule allows a gene to influence another either directly or indirectly via other genes, making it challenging to discern direct regulations solely from gene expression data.

However, the effect of trajectory is almost tied in the GENIE3 model.
Although augmented trajectories may be helpful in some instances, the formulation of GENIE3 is not designed to leverage these trajectories.
Although both GENIE3 and dynGENIE3 train random forest models and extract the network in the same way, GENIE3 aims to capture interactions from static states, while dynGENIE3 focuses on dynamic modeling.
Thus, the performance of GENIE3 with dynamic information is not significantly better than that without trajectories.
Additionally, data augmentation introduces noise to GENIE3, which may also impact performance.

\subsection{Dropout-robust measure boosts GRN inference}
\label{ssec:ablation-cell-similarity-measure}

\begin{table*}[t]
\centering
\small
\caption{AUPRC ($\times 10^{-3}$) of FlowGRN under different model settings in different experimental datasets out of 10 runs.
         ``$d_{\rm raw}$'' is the use of dropout-robust measure, and ``kNN'' is the use of shortest path distance in the kNN graph.
         The \textbf{\underline{best}} and \underline{second-best} results are highlighted in bold and underlined, respectively.
         }
\label{tab:AUPR-ablation-cell-similarity}
\begin{tabular}{lrrrrrrr}
\toprule[1pt]\midrule[0.3pt]
dataset & hESC & hHep & mDC & mESC & mHSC-E & mHSC-GM & mHSC-L \\
\midrule
\multicolumn{8}{l}{\textbf{TFs + 500 genes}} \\
 \;	FlowGRN  		          & 12.526$\pm$0.112 		& \underline{\textbf{12.660$\pm$0.186}} 		& \underline{\textbf{2.315$\pm$0.030}} 		& 1.826$\pm$0.152 		& \underline{\textbf{3.364$\pm$0.233}} 		& \underline{\textbf{3.333$\pm$0.145}} 		& \underline{\textbf{2.348$\pm$0.139}} \\
    \; \; W/o kNN 		& \underline{\textbf{12.541$\pm$0.150}} 		& \underline{12.644$\pm$0.179} 		& 2.302$\pm$0.030 		& \underline{1.835$\pm$0.154} 		& \underline{3.354$\pm$0.247} 		& 3.268$\pm$0.081 		& 2.292$\pm$0.161 \\
    \; \; W/o $d_{\rm raw}$ 		& 12.536$\pm$0.105 		& 12.628$\pm$0.172 		& 2.306$\pm$0.027 		& \underline{\textbf{1.858$\pm$0.126}} 		& 3.306$\pm$0.342 		& 3.295$\pm$0.091 		& 2.254$\pm$0.144 \\
 	\; \; W/o $d_{\rm raw}$, kNN & \underline{12.537$\pm$0.112} 		& 12.624$\pm$0.162 		& \underline{2.309$\pm$0.032} 		& 1.833$\pm$0.158 		& 3.293$\pm$0.176 		& \underline{3.312$\pm$0.089} 		& \underline{2.304$\pm$0.159} \\
\midrule
\multicolumn{8}{l}{\textbf{TFs + 1,000 genes}} \\
    \; FlowGRN  		          	& 8.815$\pm$0.147 		& \textbf{\underline{8.978$\pm$0.761}} 		& 1.270$\pm$0.024 		& \underline{0.808$\pm$0.048} 		& \textbf{\underline{1.602$\pm$0.109}} 		& 1.556$\pm$0.067 		& \underline{0.931$\pm$0.041} \\

  \; \; W/o kNN 			& 8.817$\pm$0.144 		& \underline{8.907$\pm$0.759} 		& \textbf{\underline{1.282$\pm$0.026}} 		& \textbf{\underline{0.811$\pm$0.042}} 		& 1.540$\pm$0.065 		& \textbf{\underline{1.592$\pm$0.097}} 		& \textbf{\underline{0.933$\pm$0.041}} \\
    \; \; W/o $d_{\rm raw}$ 			& \underline{8.845$\pm$0.140} 		& 8.889$\pm$0.646 		& \underline{1.278$\pm$0.030} 		& 0.805$\pm$0.043 		& \underline{1.587$\pm$0.093} 		& \underline{1.586$\pm$0.100} 		& 0.921$\pm$0.037 \\
 	\; \; W/o $d_{\rm raw}$, kNN	& \textbf{\underline{8.845$\pm$0.138}} 		& 8.883$\pm$0.635 		& 1.277$\pm$0.025 		& 0.802$\pm$0.044 		& 1.575$\pm$0.053 		& 1.557$\pm$0.077 		& 0.916$\pm$0.024 \\
\midrule[0.3pt]\bottomrule[1pt]
\end{tabular}
\end{table*}

We conducted another ablation study to justify the design of our dropout-robust measure on the experimental datasets.
We compared the FlowGRN model with and without the dropout-robust measure, and with and without the kNN graph geodesic distance.
Both components are defined in Section~\ref{ssec:dropout-robust-cell-similarity}, and the results are shown in Table~\ref{tab:AUPR-ablation-cell-similarity}.
Although performance differences are minor and within standard deviations, we can see that the dropout-robust measure $d$ is the primary driver of performance improvement, especially on the ``TFs + 1,000 genes'' dataset.
This shows that the dropout-robust measure is essential for the GRN inference task, as it helps to avoid the noise introduced by dropout.

\subsection{Visualization of reconstructed trajectories}

As part of the quality check, we visualized the reconstructed trajectories $T$ of the 7 experimental datasets with two gene set sizes in Figures~\ref{fig:traj_qc_500}--\ref{fig:traj_qc_1000} in the Appendix.

In most datasets, the reconstructed trajectories broadly follow the progression of the observed cell distributions in the PCA space, although some trajectories leave dense cell regions or bend sharply.
A clear example is the hHep-1000 dataset in Figure~\ref{fig:traj_qc_1000}, where several reconstructed trajectories twist around $t=4$.
This pattern may reflect the rapid change in cell distributions between $t=3$ and $t=5$, although the 2-dimensional PCA projection may also introduce apparent overlaps that are absent from the original space.
Therefore, these visualizations provide a qualitative sanity check rather than evidence of accurate cell-level trajectory reconstruction.

In addition, we observed that the CFM model also gives velocity predictions outside the observed cell distributions, and these non-zero velocity predictions potentially misguide trajectory reconstruction during integration, once the reconstructed trajectory leaves the observed cell distributions.
This observation further reminds us that the learned vector field is only valid in the observed cell distributions.

\section{Conclusion and Limitations}
\label{sec:conclusion}
In this paper, we proposed FlowGRN, a novel deep learning model for inferring GRNs from scRNA-seq data.
FlowGRN is a combination of a CFM model and an ensemble tree model that reconstructs gene expression trajectories and infers the GRN from the reconstructed trajectories.
To mitigate the impact of dropout, we proposed a dropout-robust cell similarity measure that captures cell-cell similarity in the presence of dropout.

We evaluated FlowGRN on the BEELINE benchmark and compared it with eight state-of-the-art GRN inference models.
FlowGRN demonstrated top-tier performance in both synthetic and curated datasets, achieving the best performance in 6 of 10 tasks.
It has advantages in learning regulation types and modeling dynamics with long cascades and cycles, which are well-known challenges in GRN inference.
FlowGRN is also scalable to high-dimensional experimental datasets of up to 1,783 genes with dropout, thanks to the advanced CFM model and the dropout-robust cell similarity measure.
We also conducted ablation studies to validate the design of FlowGRN, demonstrating that the dropout-robust cell similarity measure and trajectory reconstruction are essential for our model.

FlowGRN has some limitations.
First, FlowGRN exhibits a notable performance drop on datasets with few cells, such as mDC and mESC, which is a common pitfall of deep learning models.
Second, our method cannot model the dynamics and infer network structure in a unified manner, which limits its interpretability.
In the future, we will combine dynamic modeling and GRN inference in a unified framework by restricting the CFM model with biological constraints, following the approach of physics-informed neural networks (PINN).
We will also replace the Euclidean distance in the CFM training with our similarity measure and align the trajectories with the tendency of the cell distribution shifts.
We expect that these improvements will further enhance the performance of FlowGRN on high-dimensional datasets.

\section*{Conflicts of Interest}
The authors declare no conflicts of interest.

\section*{Data and Code Availability}
We do not generate any new data in this work.
All the data used in this work are publicly available.
The code of FlowGRN is available at \url{https://github.com/1250326/FlowGRN}, implemented in Python 3.9.

\begin{acks}
Authors Tsz Pan Tong and Jun Pang acknowledge financial support from the Institute for Advanced Studies of the University of Luxembourg through an Audacity Grant (AUDACITY-2021).
This work was also supported by the Luxembourg National Research Fund (FNR) under the grant agreement INTER/NCN/24/18732364/EdgeCR.
\end{acks}

\bibliographystyle{ACM-Reference-Format}
\bibliography{reference}

\newpage
\appendix

\section{Appendix}
\label{sec:appendix}

\subsection{Network statistics and baseline descriptions}
\label{ssec:network-statistics}

Table~\ref{tab:syn-cur-network-statistics} shows the network statistics of the synthetic and curated datasets in the BEELINE benchmark.
The number of genes in the dataset, the number of edges in the ground truth GRN, and the network density are reported, where the network density is defined as $\# \text{Edges}/g^2$.

\begin{table}[!h]
\centering
\caption{Network Statistics of the Synthetic and Curated Datasets}
\label{tab:syn-cur-network-statistics}
\begin{tabular}{lrrr}
\toprule[1pt]\midrule[0.3pt]
Dataset & \# Genes & \# Edges & Network Density \\
\hline
\multicolumn{4}{l}{\textbf{Synthetic}} \\
\quad  BF      & 7        & 12       & 0.24            \\
\quad  BFC     & 10       & 18       & 0.18            \\
\quad  LI      & 7        & 8        & 0.16            \\
\quad  LL      & 18       & 19       & 0.06            \\
\quad  CY      & 6        & 6        & 0.17            \\
\quad  TF      & 8        & 20       & 0.31            \\
\hline
\multicolumn{4}{l}{\textbf{Curated}} \\
\quad  mCAD    & 5        & 14       & 0.56            \\
\quad  VSC     & 8        & 15       & 0.23            \\
\quad  HSC     & 11       & 30       & 0.25            \\
\quad  GSD     & 19       & 79       & 0.22            \\
\midrule[0.3pt]\bottomrule[1pt]
\end{tabular}
\end{table}

Table~\ref{tab:models_compare} compares FlowGRN with BEELINE baseline models in terms of their properties, inferred GRN types, and methodologies.
We categorize the models into 5 categories: Correlation (Corr), ODE, Regression (Reg), Mutual Information (MI), and Tree.
A model is said to be signed if it can determine the activation or inhibition of gene interactions, and the positive (negative) values in the inferred GRN correspond to the activation (inhibition).
Our FlowGRN is a combination of ODE and ensemble tree models, leveraging the advantages of both methodologies to reconstruct cell dynamics and efficiently and scalably extract gene interactions.

\begin{table*}[!ht]
    \centering
    \caption{Comparison between FlowGRN and BEELINE baseline models on their properties, inferred GRN types and methodologies.
    Model categories include Correlation (Corr), ODE, Regression (Reg), Mutual Information (MI) and Tree.
    }
    \label{tab:models_compare}
    \begin{small}
    \begin{tabular}{lccl} \toprule[1pt]\midrule[0.3pt]
                                             & Category           & ~Signed?~  & Description                                                                                          \\ \midrule  
    LEAP~\cite{specht2017leap}               & Corr        & \centering\xmark & \begin{minipage}[t]{0.55\textwidth}{A model that uses the maximum Pearson correlations at different time lags as the gene interaction strength}\end{minipage}                                                    \\[1cm]
    SCODE~\cite{matsumoto2017scode}          & ODE                & \centering\cmark & \begin{minipage}[t]{0.55\textwidth}{A linear ODE model for the gene expression level}\end{minipage}                                                                                                              \\[0.6cm]
    GRISLI~\cite{aubin2020grisli}            & ODE                & \centering\xmark & \begin{minipage}[t]{0.55\textwidth}{A model similar to SCODE but with sparsity constraint and less restrictions due to the stability selection algorithm}\end{minipage}                                           \\[1cm]
    GRNVBEM~\cite{sanchez2018grnvbem}        & Reg         & \centering\cmark & \begin{minipage}[t]{0.55\textwidth}{A multivariate AR1MA1 time series model solved by the variational Bayesian inference}\end{minipage}                                                                            \\[0.6cm]
    SINCERITIES~\cite{papili2018sincerities} & Reg         & \centering\cmark & \begin{minipage}[t]{0.55\textwidth}{Models the gene distributional shift distances by linear regression to spot significant gene interactions}\end{minipage}                                                       \\[1cm]
    Scribe~\cite{qiu2020scribe}              & MI & \centering\xmark & \begin{minipage}[t]{0.55\textwidth}{Causality inferencing model on GRN that computes restricted directed information~\cite{rahimzamani2016network}}\end{minipage} \\[1cm]
    GENIE3~\cite{huynh2010genie3}            & Tree               & \centering\xmark & \begin{minipage}[t]{0.55\textwidth}{A tree model using random forests as predictors and feature importance as gene interaction strength}\end{minipage}                                                             \\[1cm]
    GRNBOOST2~\cite{moerman2019grnboost2}    & Tree               & \centering\xmark & \begin{minipage}[t]{0.55\textwidth}{A tree model using gradient boosting machines as predictors and feature importance as gene interaction strength}\end{minipage}                                                 \\[1cm]
    FlowGRN (Ours)                             & ODE+Tree     & \centering\cmark & \begin{minipage}[t]{0.55\textwidth}{A tree model using random forests to extract gene interaction strengths from cell trajectories reconstructed by neural ODE}\end{minipage}                                                       \\ \midrule[0.3pt]\bottomrule[1pt]
    \end{tabular}
    \end{small}
\end{table*}

\subsection{Preprocessing of the experimental datasets}
\label{ssec:appendix-preprocessing}
Our FlowGRN model is designed for GRN inference from a time series of gene expression data.
However, the data in the BEELINE benchmark lack temporal information; instead, they provide pseudotime for each cell from Slingshot~\cite{street2018slingshot}.

To adapt the data for our model, we first cluster the cells by their dimension-reduced gene expression in each dataset.
We manually checked the clustering results and their alignment with the pseudotime.
Then, we used the mean pseudotime of each cluster as its sampling time.
Note that FlowGRN only requires the temporal ordering but not the exact time, so the mean pseudotime is sufficient for our model.

In implementation, we use UMAP~\cite{mcinnes2018umap} for dimension reduction.
For clustering, we use DBSCAN and KMeans, depending on the within-cluster distance after clustering.
We prioritize the use of DBSCAN, as it is more robust to noise and does not require the number of clusters to be specified in advance.
However, DBSCAN requires a hard threshold for the minimum distance between clusters, which may form large clusters with multiple cell states.
Thus, KMeans serves as a backup, and the number of clusters is set to match the manual inspection.

\subsection{Standard deviations of the reported data}
\label{ssec:additional-exp-result}

\begin{figure}[t]
  \centering
  \includegraphics[width=0.8\linewidth]{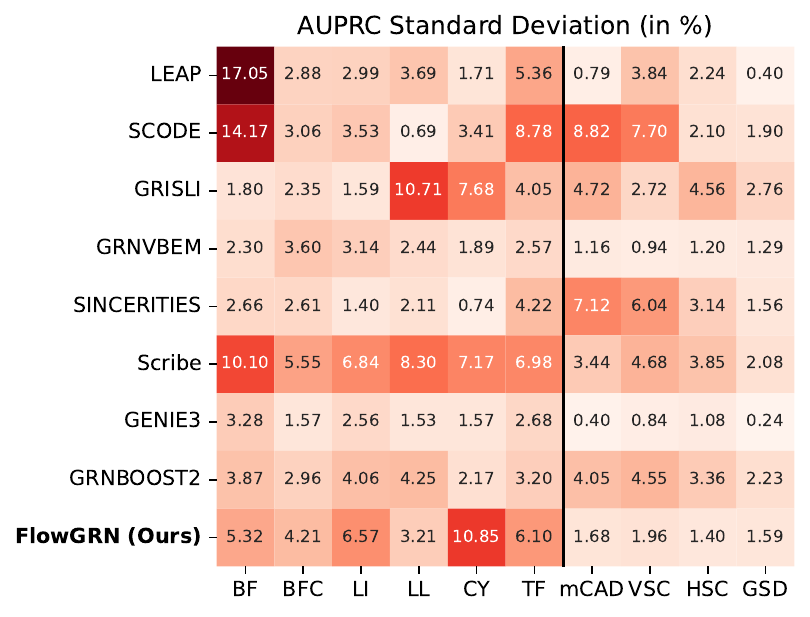}
  \Description{Standard deviation of AUPRC of directed GRN inference models on different simulated datasets.}
  \caption{Standard deviation of AUPRC of directed GRN inference models on different simulated datasets out of 10 runs.
           Synthetic and curated datasets are separated by a vertical line.}
  \label{fig:directed-std}
\end{figure}

\begin{figure}[t]
  \centering
  \includegraphics[width=0.8\linewidth]{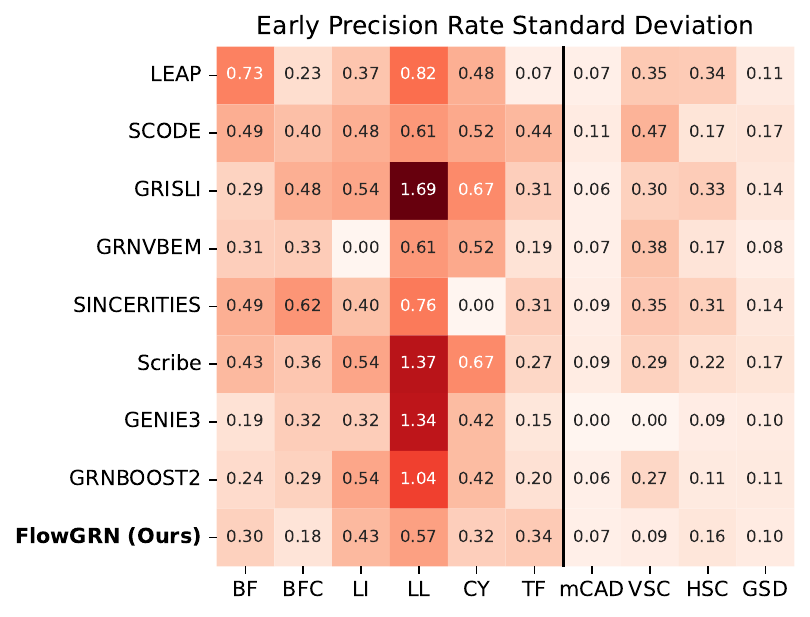}
  \Description{Standard deviation of EPR of directed GRN inference models on different simulated datasets.}
  \caption{Standard deviation of EPR of directed GRN inference models on different simulated datasets out of 10 runs.
           Synthetic and curated datasets are separated by a vertical line.}
  \label{fig:directed-epr-std}
\end{figure}

Figure~\ref{fig:directed-std} shows the standard deviation of AUPRC of the directed GRN inference models on all simulated datasets out of 10 runs.
FlowGRN has relatively high deviations in the synthetic datasets, but low deviations in the curated datasets.
Note that deterministic models (LEAP, GRISLI, GRNVBEM, SINCERITIES, and Scribe) have a nonzero standard deviation in synthetic and curated datasets because both types of datasets contain 10 independent simulations of the same GRN.
However, BEELINE only contains one scRNA-seq experiment for each experimental dataset.
Thus, the standard deviation of the deterministic models is zero.

Figure~\ref{fig:directed-epr-std} shows the standard deviation of the EPR of the directed GRN inference models on all simulated datasets out of 10 runs.
FlowGRN has an average standard deviation compared to other baselines, and the standard deviation is also low in the curated datasets.
It is notable that the standard deviations in the LL dataset are consistently high for all models, which may be attributed to the inherent difficulty of its long cascade dynamics.

\begin{figure}[t]
  \centering
  \includegraphics[width=0.8\linewidth]{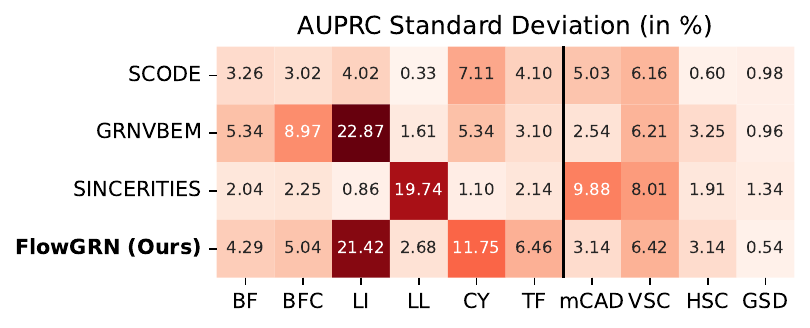}
  \Description{Standard deviation of AUPRC of directed GRN inference models with edge types on different simulated datasets.}
  \caption{Standard deviation of AUPRC of directed GRN inference models with edge types on different simulated datasets out of 10 runs.
           Synthetic and curated datasets are separated by a vertical line.}
  \label{fig:signed-std}
\end{figure}

Figure~\ref{fig:signed-std} shows the standard deviation of AUPRC of the directed GRN inference models with edge types on all simulated datasets out of 10 runs.
The standard deviation of FlowGRN is comparable to the second-best model, GRNVBEM, indicates that FlowGRN has a better performance than GRNVBEM with the same level of stability.

\subsection{Visualization of reconstructed trajectories}
\label{ssec:viz-traj}

\begin{figure}[!t]
  \centering
  \includegraphics[width=0.83\linewidth]{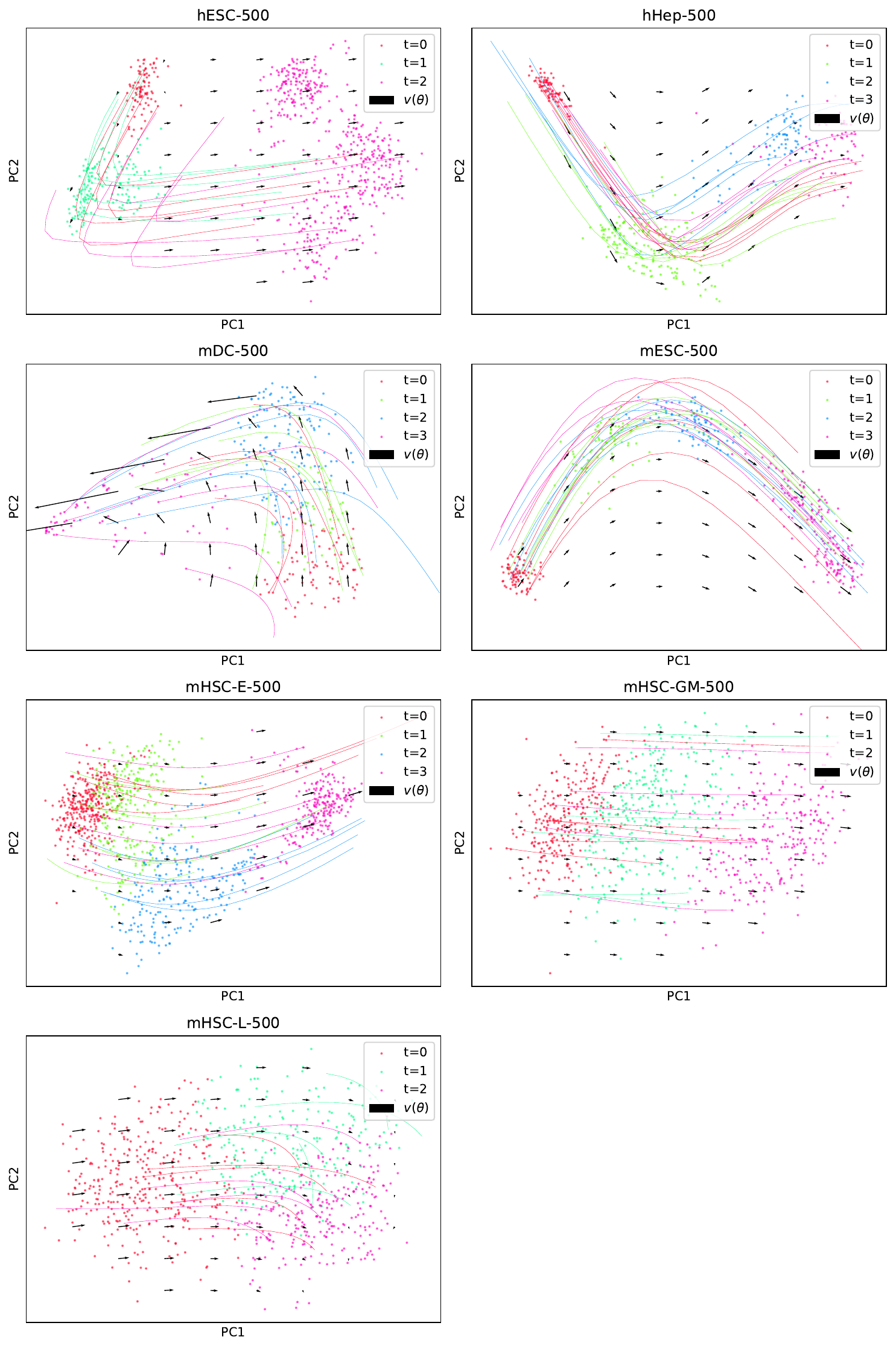}
  \Description{Visualization of the trained vector field and reconstructed trajectories of the 7 experimental ``TFs + 500 genes'' datasets.}
  \caption{Trained vector field and reconstructed trajectories of the 7 experimental ``TFs + 500 genes'' datasets.
           }
  \label{fig:traj_qc_500}
\end{figure}

\begin{figure}[!t]
  \centering
  \includegraphics[width=0.83\linewidth]{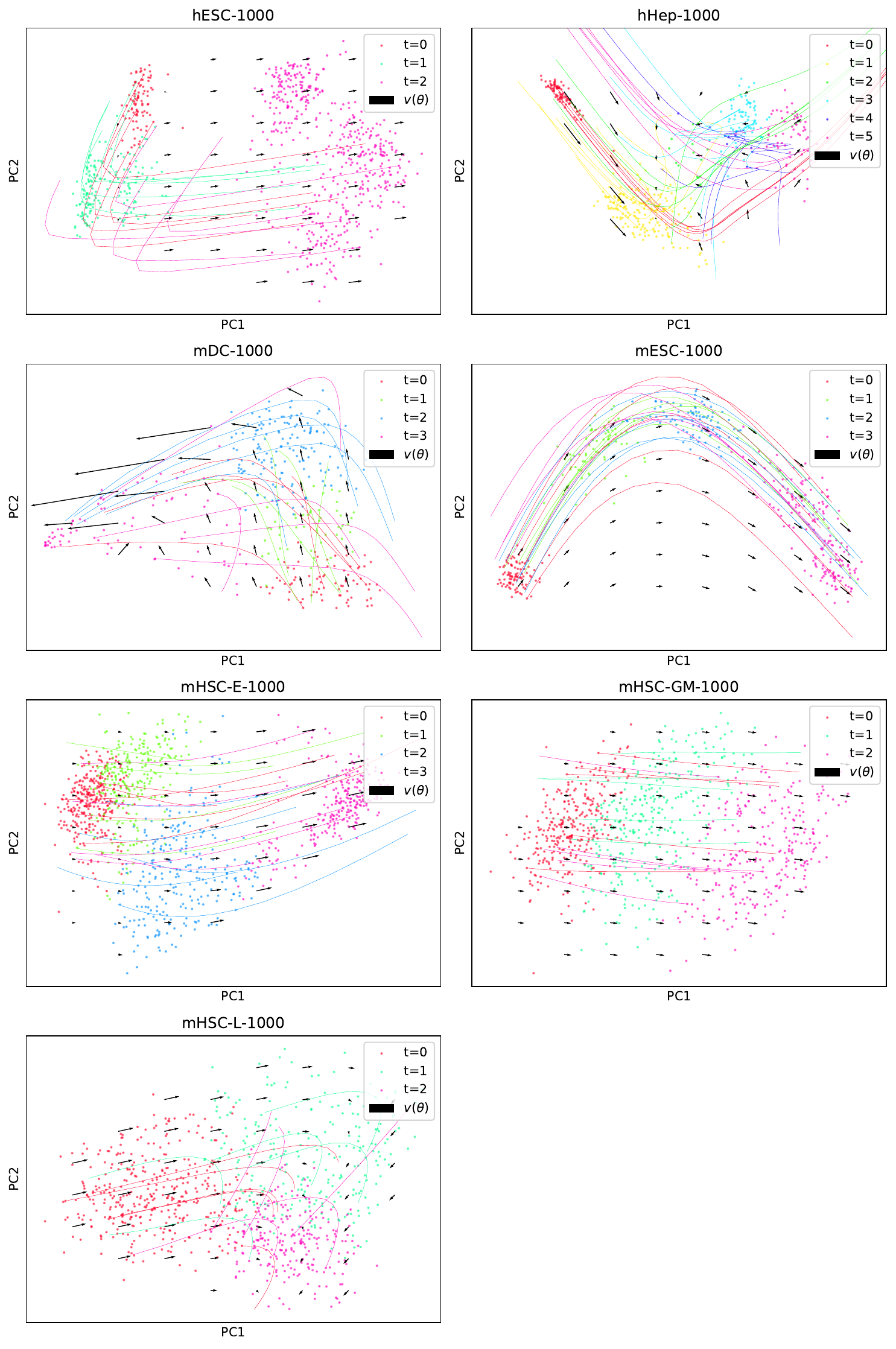}
  \Description{Visualization of the trained vector field and reconstructed trajectories of the 7 experimental ``TFs + 1,000 genes'' datasets.}
  \caption{Trained vector field and reconstructed trajectories of the 7 experimental ``TFs + 1,000 genes'' datasets.
           }
  \label{fig:traj_qc_1000}
\end{figure}

For each dataset, we use PCA to project the $g$-dimensional gene space to a 2-dimensional space.
While nonlinear dimensionality reduction methods such as UMAP~\cite{mcinnes2018umap} and t-SNE~\cite{van2008visualizing} are widely used in single-cell data visualization, they distort the reconstructed trajectories and cannot provide a continuous mapping from the original space to the low-dimensional space.
PCA learns a linear mapping $W_{pca}: \mathbb{R}^g \to \mathbb{R}^2$ to bridge the original space $\mathbb{R}^g$ and the 2-dimensional space $\mathbb{R}^2$, as well as their tangent spaces.
Denote $\mu'$ as the mean of log-transformed gene expression data, cell $x'\in \mathbb{R}^g$ is projected to $W_{pca} (x'-\mu') \in \mathbb{R}^2$, and the tangent vector $v_\theta(x', t) \in \mathbb{R}^g$ is projected to $W_{pca} v_\theta(x', t) \in \mathbb{R}^2$.
Since the projection is linear, the projected trajectories and vector field remain continuous, although the projection may distort distances and cause distinct trajectories to overlap.

With the PCA projection, we can visualize the cells, reconstructed trajectories, and the vector field in the 2-dimensional space.
For each dataset, cell clustering and time ordering results detailed in Section~\ref{ssec:appendix-preprocessing} in the Appendix are colored in the cell scatter plot.
For each time point, we randomly sample 5 reconstructed trajectories and plot them as curves.
We also compute the vector field $v_\theta$ on a grid of points and plot it as arrows.

\end{document}